\documentclass{Latex-NSR}
\usepackage{multirow}

\begin{document}

\ArticleType{RESEARCH ARTICLE}%
\ArticleSubject{INFORMATION SCIENCE}
\Year{2020} %
\Month{? ? 2020}
\Vol{?} %
\No{?} %
\BeginPage{1} %
\EndPage{?} %
\DOI{??????????}
\AuthorMark{{\rm Zhang et al.}}

\title{Tiny noise, big mistakes: Adversarial perturbations induce errors in Brain-Computer Interface spellers}%

\author[1]{Xiao~Zhang}{}
\author[1]{Dongrui~Wu}{drwu@hust.edu.cn}
\author[2]{Lieyun~Ding}{dly@hust.edu.cn}
\author[2]{Hanbin~Luo}{}
\author[3]{Chin-Teng~Lin}{}
\author[4,5]{Tzyy-Ping~Jung}{}
\author[6]{\\ Ricardo~Chavarriaga}{}

\address[{\rm1}]{Ministry of Education Key Laboratory of Image Processing and Intelligent Control, School of Artificial Intelligence and Automation,\\ Huazhong University of Science and Technology, Wuhan 430074, China.}
\address[{\rm2}]{School of Civil Engineering and Mechanics, Huazhong University of Science and Technology, Wuhan 430074, China.}
\address[{\rm3}]{Centre of Artificial Intelligence, Faculty of Engineering and Information Technology, University of Technology Sydney, Sydney 2007,\\ Australia.}
\address[{\rm4}]{Swartz Center for Computational Neuroscience, Institute for Neural Computation, University of California San Diego (UCSD),\\ La Jolla, CA 92093, USA.}
\address[{\rm5}]{Center for Advanced Neurological Engineering, Institute of Engineering in Medicine, UCSD, La Jolla, CA 92093, USA.}
\address[{\rm6}]{ZHAW DataLab, Z\"{u}rich University of Applied Sciences, Winterthur 8401, Switzerland.}

\abstract{An electroencephalogram (EEG) based brain-computer interface (BCI) speller allows a user to input text to a computer by thought. It is particularly useful to severely disabled individuals, e.g., amyotrophic lateral sclerosis patients, who have no other effective means of communication with another person or a computer. Most studies so far focused on making EEG-based BCI spellers faster and more reliable; however, few have considered their security. This study, for the first time, shows that P300 and steady-state visual evoked potential BCI spellers are very vulnerable, i.e., they can be severely attacked by adversarial perturbations, which are too tiny to be noticed when added to EEG signals, but can mislead the spellers to spell anything the attacker wants. The consequence could range from merely user frustration to severe misdiagnosis in clinical applications. We hope our research can attract more attention to the security of EEG-based BCI spellers, and more broadly, EEG-based BCIs, which has received little attention before.\\}


\maketitle


\setlength{\baselineskip}{14pt} \setlength{\parskip}{0.5em}

\section{INTRODUCTION}

A brain-computer interface (BCI), which has been extensively used in neuroscience, neural engineering and clinical rehabilitation, offers a communication pathway that allows people to interact with computers using brain signals directly \cite{Graimann2009, Lin2017,drwuEA2020,Chavarriaga2010}. There are many approaches to collecting signals from the brain. Electroencephalogram (EEG), usually measured from the scalp, may be the most popular one due to its simplicity and low cost \cite{Nicolas-Alonso2012}.

An EEG-based BCI speller allows a user to input text to a computer by thought \cite{Farwell1988,Chen2015a}. It enables people with severe disabilities, e.g., amyotrophic lateral sclerosis (ALS) patients, to communicate with computers or other people. The two main types of EEG-based BCI spellers are P300 spellers \cite{Farwell1988} and steady-state visual evoked potential (SSVEP) spellers \cite{Chen2015a}, which elicit different EEG patterns, as illustrated in Figure~\ref{fig:EEGSpellers}a.

A P300 speller, which uses P300 evoked potentials as its input signal \cite{Sutton1965}, was first invented by Farwell and Donchin in 1988 \cite{Farwell1988} and further developed by many others \cite{Donchin2000, Meinicke2003, Xu2004, Guan2004}. P300 is a positive deflection in voltage, typically appearing around 250 to 500 ms after a rare target stimulus occurs \cite{Polich2007}. It is an endogenous potential linked to people's cognitive processes, such as information processing and decision making \cite{Chapman1964, Sutton1967}. The standard oddball paradigm is usually used to elicit P300, in which rare target stimuli are mixed with high-probability non-target ones. The P300 speller considered in this article uses a $6 \times 6$ character matrix, which consists of 26 letters and 10 other symbols, as shown in Figure~1b. The user stares at the character he/she wants to input, while a row or column is rapidly intensified sequentially. The corresponding EEG signals are recorded and classified as a target (containing P300) or non-target (not containing P300) for each intensification. Then, the computer identifies the character at the intersection of the target row and the target column, which elicit the largest P300s, as the output. For reliable performance, each row and column may have to be intensified multiple times, which reduces the speed of the P300 speller.

Compared with the P300 speller, an SSVEP speller has the advantages of high information transfer rate (ITR), little user training, and some immunity to artifacts \cite{Beverina2003,Wang2008,Vialatte2010}. When the user stares at a visual target flickering at a specific frequency, usually between 3.5~Hz and 75~Hz, electrical signals of the same frequency, as well as its corresponding harmonics, can be observed from the EEG signals \cite{Beverina2003}. In an SSVEP speller, the pictures of different characters are flickering at different frequencies, so that a classifier can directly identify the output character from a large number of candidates by matching their flickering frequencies with the user's EEG oscillation frequency. Since all characters in an SSVEP speller are flickering simultaneously (in contrast to sequential intensification in a P300 speller), they can have much higher ITRs. The SSVEP speller considered in this study has 40 characters (Figure~1c), whose stimulation frequencies are from 8~Hz to 15.8~Hz with 0.2~Hz increment \cite{Wang2017}.

Machine learning is used in BCI spellers to construct the classifiers to detect the brain responses to stimuli (i.e. the P300 or SSVEPs patterns). Most studies so far focused on making the BCI classifiers faster and more reliable; however, few have considered their security. It has been found in other application domains that adversarial examples \cite{Szegedy2014}, which are normal examples contaminated by deliberately designed tiny perturbations, can easily fool machine learning models. These perturbations are usually so small that they are indistinguishable to human eyes. Existing studies on adversarial examples focused largely on deep learning models for computer vision. For example, it was found that a picture of a panda, after adding a weak adversarial perturbation, can be misclassified as a gibbon by a deep learning classifier \cite{Goodfellow2015}. Kurakin \emph{et al.} \cite{Kurakin2017} found that printed photos of adversarial examples can degrade the performance of an ImageNet Inception classifier. Athalye \emph{et al.} \cite{Athalye2018} 3D printed a turtle with an adversarial texture, which was classified as a riffle from almost every viewpoint. Recently, adversarial examples were also found in traditional machine-learning models \cite{Papernot2016} and many other application domains, e.g., speech recognition \cite{Carlini2018}, text classification \cite{Jia2017}, malware identification \cite{Grosse2016}, etc. Due to the high risk of adversarial attacks, many defense mechanisms have been proposed, such as defensive distillation \cite{Papernot2016a}, adversarial training \cite{Goodfellow2015,Madry2018,Tramer2018}, and so on \cite{Samangouei2018,Xie2018,Qin2019}. However, these approaches only improve empirical adversarial robustness, which is not certified and may be broken by a stronger attack approach \cite{Carlini2017,Athalye2018a}. Recently, researchers started to investigate provable guarantees of adversarial robustness, yet there is still a huge gap between certified robustness and empirical robustness \cite{Gowal2018,Cohen2019,Li2019,Balunovic2020}.

This article aims to expose a critical security concern in EEG-based BCI spellers, and more broadly, EEG-based BCIs, which has received little attention before. It shows for the first time that one can generate tiny adversarial EEG perturbation templates for target attacks for both P300 and SSVEP spellers, i.e., mislead the classification to any character the attacker wants, regardless of what the user intended character is. 
The consequence could range from merely user frustration to severe misdiagnosis in clinical applications \cite{Zhang2019}. We believe a new and more detailed understanding of how adversarial EEG perturbations affect BCI classification can inform the design of BCIs to defend against such attacks.

There have been some studies on adversarial attacks of time-series signals \cite{Carlini2018,Fawaz2019,Zhang2019,Qin2019a}. They treated time-series signals just as images, and then applied essentially the same attack approaches in image classification to generate adversarial perturbations. As a result, they need to know the full time-series before computing the adversarial perturbations, which means these approaches are not causal and hence cannot be implemented in real-world applications. For example, to attack a voice command, previous approaches need to record the entire voice command first, and then design the perturbation. However, once the perturbation is obtained, the voice command has already been sent out (e.g., to a smartphone or Amazon Echo), so there is no chance to add the perturbation to the voice command to actually perform the attack.

What distinguishes the attack approaches in this article most from previous ones is that it explicitly considers the causality in designing the perturbations. The adversarial perturbation template is constructed directly from the training set and then fixed. So, there is no need to know the test EEG trial and compute the perturbation specifically for it. The perturbation can be directly added to a test EEG trial as soon as it starts, hence satisfies causality and can be implemented in practice. Thus, it calls for an urgent need to be aware of such attacks and defend against them.

A closely related concept is universal adversarial perturbations \cite{Moosavi-Dezfooli2017}, which can also be viewed as adversarial perturbation templates and have been used to attack deep learning models in image classification. This study focuses on the security of a traditional and most frequently used BCI pipeline, which consists of separate feature extraction and classification steps, whereas universal adversarial perturbations are usually designed for non-target attacks of end-to-end deep learning models.

To summarize, our contributions are:
\begin{enumerate}
  \item We show, for the first time, that tiny noise can significantly manipulate the outputs of P300 and SSVEP spellers, exposing a critical security concern in BCIs.
  \item Instead of Deep Learning models, we consider the classical BCI pipeline consisting of feature extraction and classification as our victim models, which dominate practical BCI spellers.
  \item Our generated adversarial perturbation templates satisfy the causality of time-series signals, which rarely drew much attention before.
\end{enumerate}

\section{RESULTS} \label{sect:Results}

\subsection{Performance evaluation}

We used two measures to evaluate the performance of a BCI speller: The classification accuracy and the ITR \cite{Wolpaw1998}, which measures the typing speed of the speller:
\begin{align}
\mathrm{ITR}=\frac{1}{T}\left[\log_2Q+R\log_2R+(1-R)\log_2\frac{1-R}{Q-1}\right],\label{eq:ITR}
\end{align}
where $T$ is the average time (minutes) spent to input a user character, $Q$ the number of different characters (which was 36 in our P300 speller and 40 in the SSVEP speller), and $R$ the classification accuracy. The unit of ITR is bits/min. When the classification accuracy is lower than a random guess, i.e., $R\leq\frac{1}{Q}$, the ITR is directly set to 0.

To distinguish between the character the user wants to spell, and the one the attacker wants to mislead to, we denote the former \emph{user character}, and the latter \emph{attacker character}. Accordingly, \emph{user score} and \emph{user ITR} are used to describe the classification accuracy of user characters and the corresponding ITR, respectively. An \emph{attacker score} is defined as the ratio that the perturbation template leads the speller to output an attacker character, and the corresponding \emph{attacker ITR} is calculated by replacing $R$ in equation (\ref{eq:ITR}) with the attacker score. A higher attacker score or attacker ITR represents a better target attack performance.

\subsection{Security of the P300 speller}

~\newline\hspace*{15pt}\textbf{Dataset}: We used a public P300 dataset (dataset II) introduced by Wolpaw \emph{et al.}~ \cite{Wolpaw2004}. It recorded 64-channel EEG signals from two subjects (A and B). The EEG data were sampled at 240~Hz, bandpass filtered to 0.1-40~Hz, then $z$-normalized for each channel. There were 85 training character trials and 100 test ones for each subject. For each trial, a set of 12 random intensifications (six rows and six columns) were repeated 15 times (i.e., each row was intensified 15 times, and each column was also intensified 15 times). Each intensification lasted for 100~ms, after which the character matrix was blanked for 75~ms. So, it took $(100+75)\times 12 \times 15=31,500$~ms, or 31.5~s, to input a character. The spelling speed can be improved by using fewer repeats, e.g., 10 or 5; however, the spelling accuracy generally decreases with a smaller number of repeats.

Note that all the following experiments were also successfully performed on a public ALS P300 dataset with eight ALS patients (see Supplementary Information for the details).

\textbf{The victim model}: The victim model was a Riemannian geometry based approach, which won the Kaggle BCI challenge\footnote{https://www.kaggle.com/c/inria-bci-challenge} in 2015. First, 16 xDAWN spatial filters \cite{Rivet2009}, eight for the target trials and another eight for the non-target trials, were designed to filter all the trials. The template-signal covariance matrices of the EEG epochs were projected onto the tangent space of a Riemannian manifold \cite{Barachant2012, Barachant2013, Yger2017}, using Affine Invariant Riemannian Metric as its distance metric. Finally, we classified the feature vectors with a Logistic Regression model in the tangent space. The details can be found in the Supplementary Information. The model was trained with class-specific weights to accommodate class imbalance. All operations in these blocks are differentiable, so we re-implemented them using Tensorflow \cite{Abadi2016} to facilitate the gradient calculation.

To get the label (target or non-target) of an intensification, an epoch between 0-600~ms from the beginning of the intensification was extracted and fed into the victim model to calculate the target probability. Because each row and column was intensified multiple times, voting was performed for each trial to get the target row and target column, and hence the target character.

\textbf{Baseline performance}: The first part of Table~\ref{tab:Tab1} shows the baseline performance of the clean EEG data (without adding any perturbations). As the number of intensification repeats increased, the user score increased, indicating that the classification accuracy of the user characters increased. Meanwhile, the user ITR decreased, because the time needed to input each character significantly increased.

The second part of Table~\ref{tab:Tab1} shows the baseline performance when we added Gaussian noise to the raw EEG data, averaged over 10 runs. The Gaussian noise perturbations were preprocessed in the same way as the adversarial perturbations, by replacing the perturbation --$\widetilde{P}$ in equation (\ref{eq:OverallDirection})-- with standard Gaussian noise, so that they had the same energy. We use signal-to-perturbation ratio (SPR) to quantify the magnitude of the perturbation, which is also presented in the second part of Table~\ref{tab:Tab1}. Gaussian noise perturbations had almost no impact on the user score and the user ITR at all, not to mention forcing the P300 speller to output a specific attacker character. These results suggest that more sophisticated adversarial perturbations are needed to attack the P300 speller.

\textbf{Performance under adversarial attacks}: Then, we added the adversarial perturbation template to the test EEG trials to validate whether it was effective in misleading the P300 speller. Figure~\ref{fig:P300Analysis}a shows the attacker scores of the 36 characters. The attacker can manipulate the P300 speller to spell whatever character he/she wants, regardless of what the user intended character is, with a higher than 90\% average success rate.

The third part of Table~\ref{tab:Tab1} shows the average user scores and ITRs with different numbers of intensification repeats. The user scores and ITRs were close to zero, suggesting that the user almost cannot correctly input the character he/she wanted.

The fourth part of Table~\ref{tab:Tab1} shows the average attacker scores and ITRs with different numbers of intensification repeats. The attacker score increased with the number of intensification repeats, because more repeats increased the number of times that the attacker can inject the perturbation into the benign EEG trial.

To better quantify the magnitude of the perturbations, we also calculated two SPRs. The adversarial perturbation template was only added at some specific periods of the EEG trial, as shown in Figure~\ref{fig:P300Analysis}b, therefore we defined a \textit{period SPR} to measure the SPR of the perturbed period, and also a \textit{trial SPR} to measure the SPR of the entire trial. The last part of Table~\ref{tab:Tab1} shows these SPRs. They were higher than 20~dB, suggesting that the adversarial perturbation template may be undetectable when added to benign EEG trials.

\textbf{Visualization of the adversarial perturbations}: In addition to high attack performance, another requirement in adversarial attacks is that the perturbations should not be detected easily.

Figure~\ref{fig:P300Analysis}b shows a typical EEG trial before and after the adversarial perturbation on Subject~A. For clarity, we only show channels F3, F4, Cz, P3 and P4, which evenly distribute on the scalp. One can barely distinguish the adversarial EEG trial from the original EEG trial.

A traditional way to visualize the P300 signal is to take the average of multiple P300 trials. We also took this approach to check if there was a noticeable difference between the average target (or non-target) trials, before and after perturbation. Figure~\ref{fig:P300Analysis}c shows the results from the Cz channel. One can hardly observe any differences. Figure~\ref{fig:P300Analysis}c also shows the spectrograms and topoplots of the difference between the average target EEG trial and the average non-target EEG trial. The original and adversarial spectrograms (or topoplots) show very similar energy distributions, and are hardly distinguishable by human eyes.

\subsection{Security of the SSVEP Speller} \label{sect:AttackOnSSVEP}

\begin{figure}[ht]         \centering
    \includegraphics[width=\linewidth,clip]{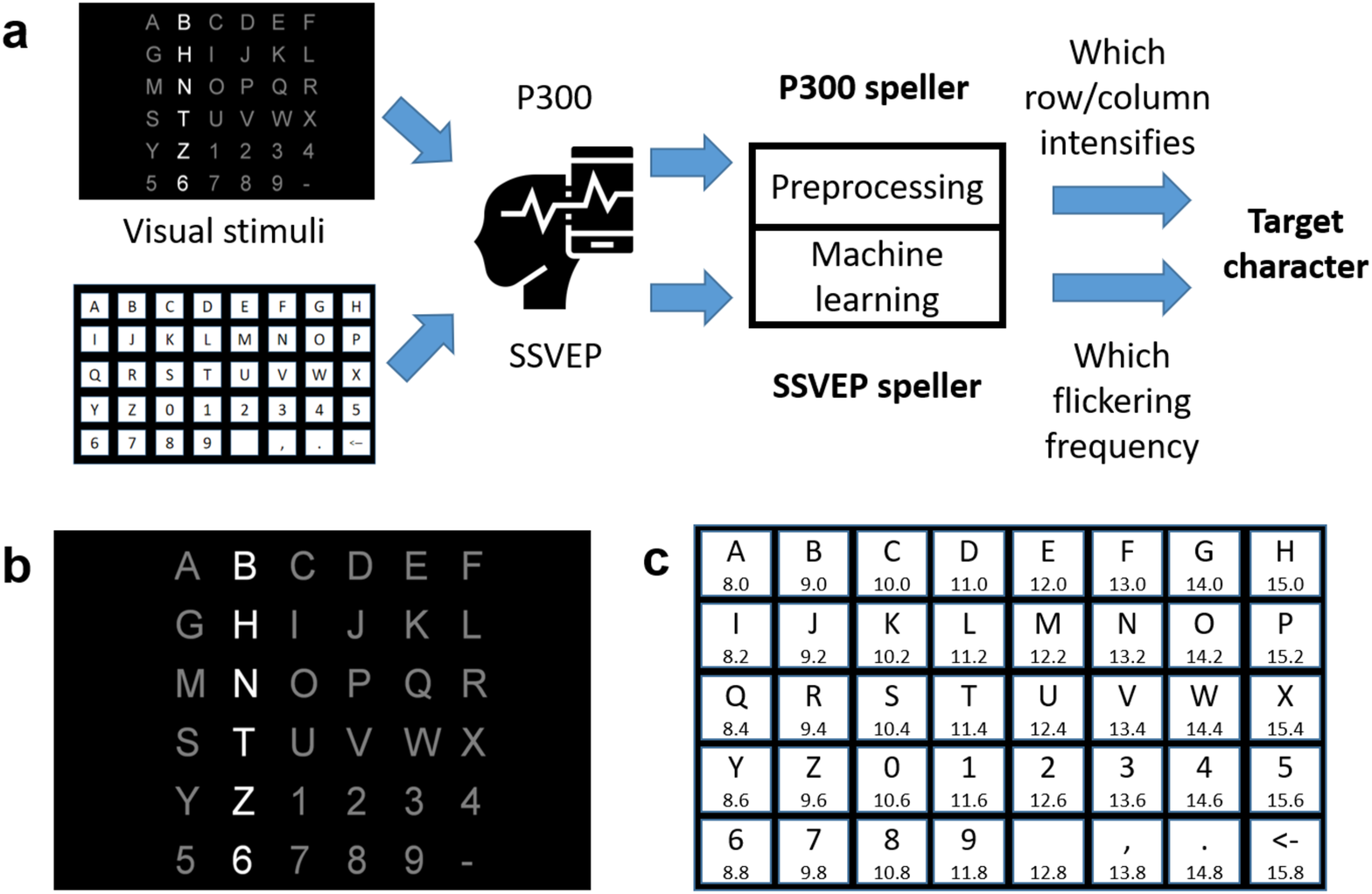}
\caption{A P300 speller and an SSVEP speller. \textbf{a}, Workflow of a P300 speller (top path) and an SSVEP speller (bottom path). For each speller, the user watches the stimulation interface, focusing on the character he/she wants to input, and EEG signals are recorded and analyzed by the speller. The P300 speller first identifies the row and the column that elicit the largest P300, and then outputs the character at their intersection. The SSVEP speller identifies the output character directly by matching the user's EEG oscillation frequency with the flickering frequency of each candidate character. \textbf{b}, Stimulation interface of a P300 speller, where the second column is intensified. \textbf{c}, Stimulation interface of an SSVEP speller. The number below each character indicates its flickering frequency (Hz).} \label{fig:EEGSpellers}
\end{figure}

\begin{figure*}[ht] \centering
    \includegraphics[width=\linewidth,clip]{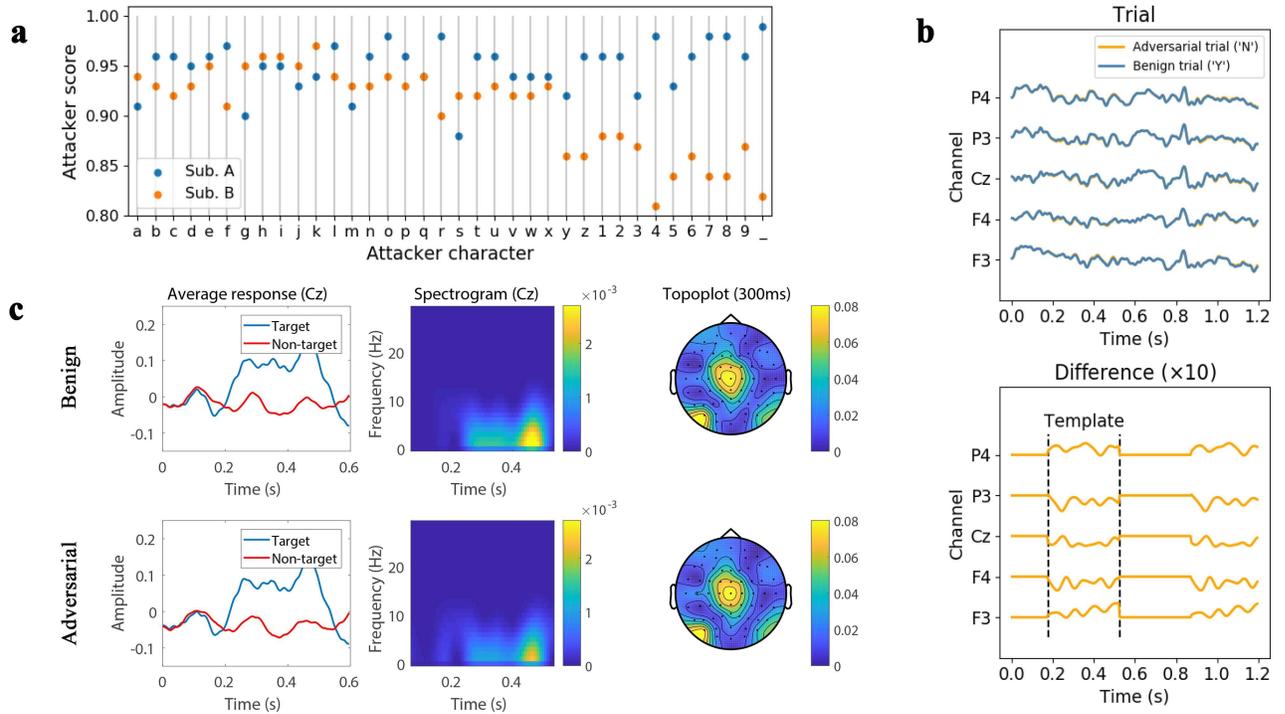}\\
\caption{P300 speller attack results. \textbf{a}, Attacker scores of manipulating the P300 speller to misclassify the 100 test character trials into a specific attacker character. The P300 speller used 15 intensification repeats for each character. \textbf{b}, EEG trials before and after adversarial perturbation, which are almost completely overlapping (the SPRs are shown in Table~\ref{tab:Tab1}), and the difference (magnified ten times) between the adversarial trial and the benign trial. The non-zero part of the difference is the adversarial perturbation template, which is added to a benign EEG trial according to the attacker character. The adversarial perturbation led the P300 speller to misclassify letter \emph{Y} into \emph{N}. \textbf{c}, left column: the average of $100\times15\times2=3,000$ target trials (containing P300) and the average of $100\times15\times10=15,000$ non-target trials (not containing P300) at channel Cz, for benign and adversarial trials; middle column: spectrogram of the difference between the average target trial and the average non-target trial in channel Cz, for benign and adversarial trials; right column: topoplot of the difference between the average target trial and the average non-target trial, for benign and adversarial trials. \textbf{b} and \textbf{c} present the visualization of the adversarial perturbations for Subject A.}\label{fig:P300Analysis}
\end{figure*}

~\newline\hspace*{15pt}\textbf{Dataset}: The dataset was first introduced by Wang \emph{et al.} \cite{Wang2017} as a benchmark dataset for SSVEP-based BCIs. The 64-channel signals were recorded from 35 subjects using an extended 10-20 system. During the experiments, the subjects were facing a monitor, in which a $5\times8$ character matrix was flickering. Different flickering frequencies were assigned to the 40 characters respectively, ranging from 8~Hz to 15.8~Hz with 0.2~Hz increment, as shown in Figure~\ref{fig:EEGSpellers}c. Six blocks of EEG signals were recorded from each subject, each with 40 trials, corresponding to the 40 target characters. Each trial was downsampled to 250~Hz and lasted 6 seconds, including 0.5~s before stimulus onset, 5~s for stimulation, and 0.5~s after stimulus offset.

Chen \emph{et al.} \cite{Chen2015} showed that an SSVEP at the stimulation frequency and its harmonics usually starts to be evoked with a delay around 130-140~ms; hence, we extracted EEG signals between [0.13, 1.38]~s after the stimulus onset as the input to the victim model. Nine channels over the occipital and parietal areas (Pz, POz, PO3, PO4, PO5, PO6, Oz, O1 and O2) were chosen. The signals were bandpass filtered to 7-90~Hz with a fourth-order Butterworth filter.

\textbf{The victim model}: Extracting the frequency information of SSVEPs is an essential step in recognizing the stimulation frequency, and hence the user character. A natural solution is to utilize fast Fourier transform to estimate the spectrum, so that the energy peaks can be matched to the stimulation frequency; however, canonical correlation analysis (CCA) was recently shown to be more promising in identifying the stimulation frequency \cite{Lin2006,Chen2015}. Thus, CCA-based frequency recognition was used in the victim model.

CCA is a statistic approach that can be used to extract the underlying correlation between two multi-channel time series \cite{Akaike1976}. Its main idea is to find a linear combination of channels for each time series, so that their correlation is maximized. When applied to SSVEP spellers, CCA is utilized to calculate the maximum correlation between the input EEG signals and a standard reference signal, which consists of the sinusoidal signal of a stimulation frequency and its $(N_q-1)$ harmonics ($N_q=5$ in our case).

Mathematically, let $X\in\mathbb{R}^{N_e \times N_s}$ denote an EEG trial with $N_e$ channels and $N_s$ samples, and $Y_f$ a standard reference signal of stimulation frequency $f$. The $(c,n)$-th entry of $Y_f$ is:
\begin{align}
Y_f(c,n) =
    \begin{cases}
    \sin\left((c+1) \pi \frac{f}{f_s} n\right),& \text{$c$ is odd}\\
    \cos\left(c \pi \frac{f}{f_s} n\right),& \text{$c$ is even}
    \end{cases},
\end{align}
where $f_s$ is the sampling rate, $1\leq c\leq2N_q$, and $1\leq n\leq N_s$. To calculate the maximum correlation coefficient $\rho(X, Y_f)$, $X$ and $Y_f$ are first $z$-normalized, and then $\rho(X, Y_f)$ is computed as the square root of the largest eigenvalue of matrix
\begin{align}
S(X,Y_f)=(XX^T)^{-1}XY_f^T(Y_fY_f^T)^{-1}Y_fX^T, \label{eq:Sxy}
\end{align}
i.e.,
\begin{align}
\rho(X, Y_f) = \sqrt{\lambda_{\max}\left(S(X,Y_f)\right)}. \label{eq:rho}
\end{align}
More detailed derivations can be found in the Supplementary Information.

Let $F=\{f_i\}_{i=1}^K$ be the set of $K$ candidate stimulation frequencies ($K=40$ in our case). Then, the SSVEP speller outputs the character corresponding to the following stimulation frequency:
\begin{align}
f^* = \arg\max_{f\in F}\rho(X, Y_f).
\end{align}

\textbf{Baseline performance}: Among the 35 subjects, eight with the best baseline performances (shown in the first part of Table~\ref{tab:Tab2}) were used in our experiments.

Because SSVEPs are highly susceptible to periodic noise, we evaluated the robustness of the victim model to Gaussian noise and sinusoidal noise of a random single frequency chosen from 40 stimulation frequencies, and a random phase chosen from $-\frac{\pi}{2}$ to $\frac{\pi}{2}$. We also considered compound sinusoidal noise, which can be regarded as the summation of single sinusoidal noise of different frequencies, random amplitudes, and random phases. The SPRs were all set to 25 dB, so that the energy of the Gaussian noise and single/compound (S/C) periodic noise was comparable to that of the adversarial perturbation templates. The `Gaussian Noise' and `S/C Periodic Noise' panels of Table~\ref{tab:Tab2} show the results on these noisy data, averaged over 10 runs, respectively. The victim model was almost completely immune to the Gaussian noise. The single periodic noise degraded the model performance more than the Gaussian noise or compound periodic noise.

\begin{figure}[ht]\centering
    \includegraphics[width=\linewidth,clip]{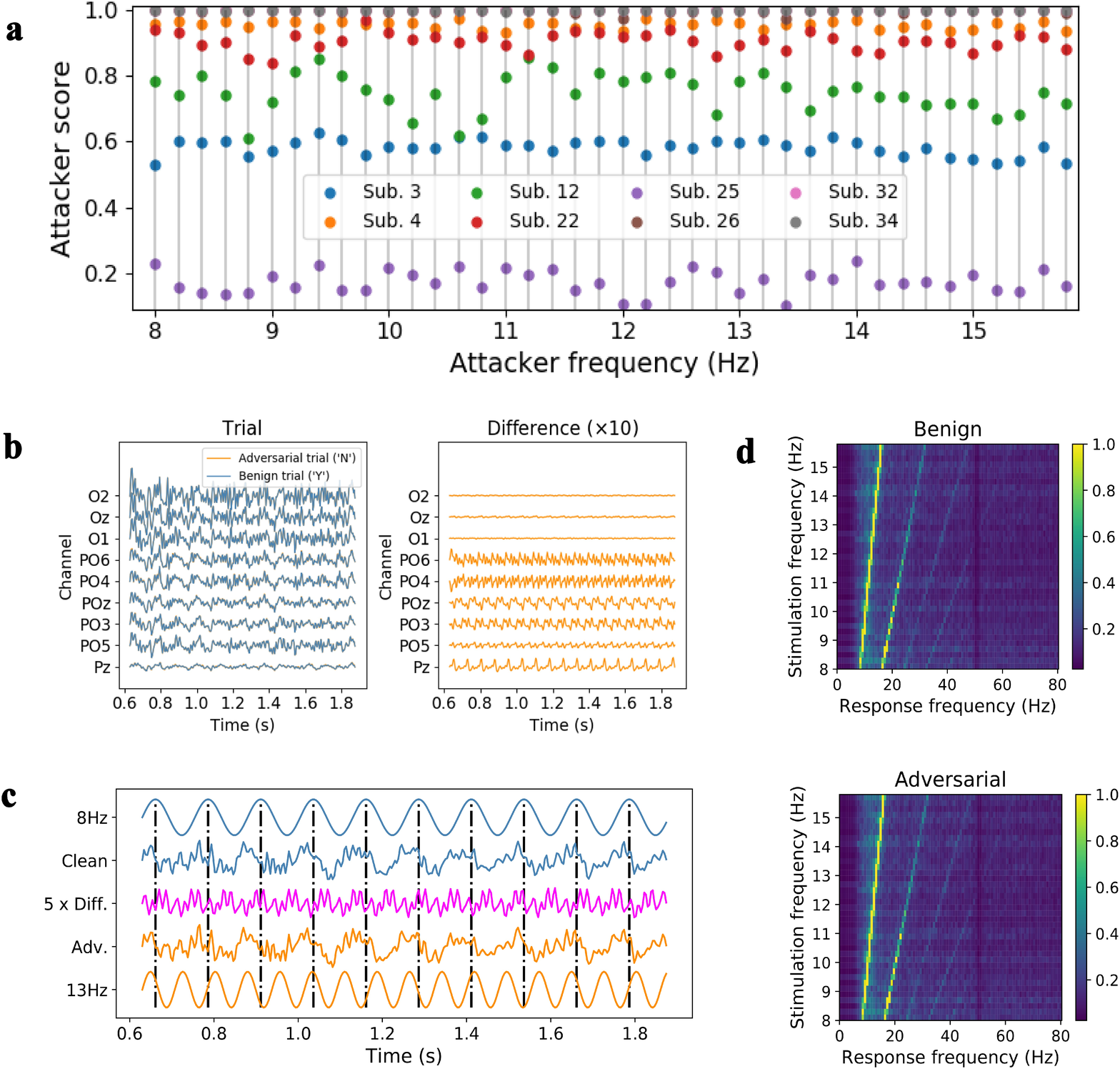} \\
\caption{SSVEP speller attack results. \textbf{a}, Attacker scores of manipulating the SSVEP speller to misclassify the $5\times40=200$ test character trials into a specific attacker frequency (character). \textbf{b}, left column: EEG trials before and after adversarial perturbation, for Subject~26; right column: the difference (adversarial perturbation) between the adversarial EEG trial and the benign EEG trial for Subject~26, magnified by ten times to make them visible. The adversarial perturbation led the SSVEP speller to misclassify the letter \emph{Y} into \emph{N}. \textbf{c}, detailed signal analysis for Channel \emph{POz} of Subject~26. The clean signal was the average of all six trials of 8~Hz stimulation frequency, and the adversarial trial was the average of the same trials with $\delta_{\text{13Hz}}$ added. Standard 8~Hz and 13~Hz sinusoidal signals are shown as references. The green dot-dashed lines mark the 8~Hz periodicity. \textbf{d}, Normalized spectra of SSVEPs for 40 stimulation frequencies, averaged over all the chosen channels and all 40 subjects.} \label{fig:SSVEPAnalysis}
\end{figure}

\textbf{Performance under adversarial attacks}: We generated 40 adversarial perturbation templates, each forcing the SSVEP speller to output a specific character. Figure~\ref{fig:SSVEPAnalysis}a shows their attacker scores. For six of the eight subjects, their output character can be manipulated to any character the attacker wanted, at 70\%-100\% success rate. Interestingly, due to individual differences, Subjects 3 and 25 showed some resistance to adversarial perturbation templates.

The fifth and sixth parts of Table~\ref{tab:Tab2} show the averaged user and attacker performances, respectively. The adversarial perturbation templates were very effective on most subjects (except Subjects 3 and 25), reducing both the user scores and the user ITRs to almost zero, i.e., the user almost cannot correctly input any character he/she wanted. The attacker scores for five subjects were close to one, i.e., the attacker was able to force the SSVEP speller to output any character he/she wanted. The SPRs were all around 25 dB, comparable to the SPRs for random noise.

\textbf{Visualization of the adversarial perturbations}: This subsection shows the characteristics of the adversarial perturbation templates, and verifies their imperceptibility to some widely-used approaches for evaluating the quality of SSVEPs.

Figure~\ref{fig:SSVEPAnalysis}b shows the EEG signals before and after adversarial perturbations, along with the magnified difference. The SSVEP speller misclassified the user character, which was supposed to be \emph{Y} (8.6~Hz), into \emph{N} (13.2~Hz). Human eyes can barely recognize the difference between the benign and the adversarial EEG trials. After being magnified by 10 times, the perturbation looks periodical, which can modify the user frequency to the attacker frequency.

We compared the clean and adversarial EEG signals with standard sinusoidal signals in Figure~\ref{fig:SSVEPAnalysis}c, using Subject~26 as an example. We took the average of the clean temporal waveforms of 8~Hz SSVEPs from Channel \emph{POz}, and did the same for their adversarial signals with $\delta_{\text{13Hz}}$ added (which forced the SSVEP speller to output the character of 13~Hz stimulation frequency). We chose Channel \emph{POz} because the adversarial perturbation on this channel had one of the largest amplitudes, as shown in Figure~\ref{fig:SSVEPAnalysis}b. Figure~\ref{fig:SSVEPAnalysis}c shows that both clean and adversarial EEG signals were synchronized with the standard 8~Hz sinusoidal signal, indicated by the green dot-dashed lines. Comparing the 13~Hz sinusoidal signal with the magnified difference, the synchronization can also be observed, suggesting that the adversarial perturbation template introduced a frequency component matching the attacker character, which was imperceptible to human eyes but powerful enough to mislead the SSVEP speller.

Figure~\ref{fig:SSVEPAnalysis}d shows the spectrum analysis of SSVEPs for 40 stimulation frequencies. We averaged the spectra of the benign EEG signals of the same stimulation frequency from all the subjects and all chosen channels, so that background activities can be suppressed. The first row of Figure~\ref{fig:SSVEPAnalysis}d, for benign trials, clearly shows that the visual stimulus, flickering at a stimulation frequency, can evoke SSVEPs of the same frequency and its harmonics. The second row of Figure~\ref{fig:SSVEPAnalysis}d shows the same property of adversarial trials, whose attacker character was randomly chosen and fixed for each stimulation frequency. We cannot observe noticeable differences between the two rows in Figure~\ref{fig:SSVEPAnalysis}d, demonstrating the challenge in detecting the adversarial perturbation templates.

\section{CONCLUSION AND DISCUSSION} \label{sect:Conclusion}

This article shows that one can generate adversarial EEG perturbation templates for target attacks for both P300 and SSVEP spellers, i.e., deliberately-designed tiny perturbations can manipulate an EEG-based BCI speller to output anything the attacker wants with high success rate, demonstrating the vulnerability of BCI spellers. We should emphasize that the attack framework used here is not specific to the victim models used in this study. They may also be utilized to attack many other classifiers in BCIs with little modification.

\textbf{Limitations:} The current approaches have two limitations: (a) they require some subject-/model- specific EEG trials to construct the adversarial perturbation template; and, (b) they need to know the exact timing of the stimulus to achieve the best attack performance. The adversarial attacks could be more dangerous if these limitations are resolved.

The first limitation may be alleviated by utilizing the transferability of adversarial examples, which was one of the most dangerous properties of adversarial examples. It was first discovered by Szegedy \emph{et al.} \cite{Szegedy2014} in 2014 and further investigated by many others \cite{Liu2016,Papernot2016, Tramer2017,Wu2018}. The transferability means that adversarial examples generated from one model can also be used to attack another model, which may have a completely different architecture and/or be trained from a different dataset. Thus, it may be possible to construct the adversarial perturbation template from some existing subjects/models and then apply it to a new subject/model. Our Supplementary Information presents experimental results on both cross-subject and cross-model transferability of the generated adversarial perturbations.

The second limitation is that the attacker needs to know the precise time synchronization between adversarial perturbation templates and EEG signals. To study how the synchronization time delay affects the attack performance, we show the relationship between the user/attacker scores and the time delay in adding the perturbation template (see Supplementary Figure~1). It can be observed that the SSVEP perturbation template was fairly robust to the time delay whereas the P300 adversarial template was sensitive to the synchronization. For the P300 speller, when the time delay increased, the user scores increased rapidly while the attacker score decreased rapidly, suggesting that hiding the time synchronization information may help defend against adversarial attacks in the P300 spellers. However, attacks insensitive to the synchronization may also be possible. For example, the idea of \emph{adversarial patch} \cite{Brown2017}, which is a tiny picture patch that can mislead the classifier when added anywhere to a large picture to be classified, may be used to increase the robustness to the synchronization time delay. Thus, defending against the attackers may not be an easy task.

\textbf{Closed-loop BCI application considerations:} In a typical closed-loop BCI speller, the user could receive real-time feedback of his/her chosen character from the screen. If the adversarial perturbation constantly misleads the speller and returns wrong characters that do not match the user's intentional input, the user would most likely stop using the speller. The consequent may not seem serious for a user that has other means of communication; however, for patients with severe impairments that rely on BCI spellers as their sole mean of communication, e.g., ALS patients, either the attacker changes the meaning of their sentences and they cannot do anything about it, or the patients stop responding, misleading doctors/researchers into thinking they are not able to communicate at all. Both consequents can significantly impact the patients.

Although this article focused on adversarial attacks of P300 and SSVEP spellers, P300 and SSVEP are also widely used in neuro-ergonomics and assessment of cognitive states, e.g., diagnosis of patients with disorder of consciousness \cite{Li2016}. The proposed approach can be used to attack these BCI systems with little modification. The adversarial perturbation could also be a serious concern if the BCI system is used in other scenarios such as wheelchair control or exoskeleton control, where the feedback could be too late and the cost of one step mistake could be fatal. Moreover, the attacker may only start the attack in some critical conditions. The user is completely unprepared, and the consequents could be more catastrophic.

Finally, we need to emphasize again that the goal of this study is not to damage EEG-based BCIs. Instead, we aim to demonstrate that serious adversarial attacks to EEG-based BCIs are possible, and hence expose a critical security concern, which has received little attention before. Our future research will develop strategies to defend against such attacks. Meanwhile, we hope our study can attract more researchers' attention to the security of EEG-based BCIs.

\section{METHODS} \label{sect:Methods}

We detail our approaches to evaluating the vulnerability of P300 and SSVEP spellers in this section.

\subsection{Datasets and Code}

The P300 speller dataset can be downloaded from \url{http://www.bbci.de/competition/iii/\#data_set_ii} (Dataset II). The P300 speller dataset of ALS patients can be downloaded from \url{http://bnci-horizon-2020.eu/database/data-sets} (P300 speller with ALS patients (008-2014)). The SSVEP dataset can be downloaded from \url{http://bci.med.tsinghua.edu.cn/download.html}. All source code is available on GitHub (\url{https://github.com/ZhangXiao96/Speller-Attacks}).

\subsection{Attack the P300 Speller}

The main idea to construct the adversarial perturbation template was to find a universal perturbation that leads the P300 classifier to classify non-target epochs into target ones. The approach was to get the directions pointing from non-target epochs to the decision boundary of the victim model, and then sum up these directions as the universal perturbation. These directions can be identified by simply calculating the gradients of the loss with respect to the input non-target EEG epochs, assuming the decision boundary is linear. Though the victim model includes nonlinear operations, the attack approach still worked surprisingly well.

Let $X$ be an EEG trial, $y$ its label (0 for non-target, and 1 for target), $f$ the victim model which gives the label probability for each input $X$, $J(X,y,f)$ the loss function (cross-entropy loss in our case), and $D_{NT}$ the dataset containing all non-target epochs in the training set. Then, the overall direction can be computed as:
\begin{align}
    \widetilde{P} = \sum_{(X,y)\in D_{NT}}\frac{\nabla_{X}J(X,1-y,f)}{\|\nabla_{X}J(X,1-y,f)\|_{F}}. \label{eq:OverallDirection}
\end{align}

After obtaining $\widetilde{P}$, we filtered it by a fourth-order Butterworth bandpass filter of $[0.1, 15]$~Hz, extracted the first 350ms signal, and then normalized it in each channel so that the L2 norm is 1. Denote the result as $\widehat{P}$. Then, the adversarial perturbation $P$ was computed as:
\begin{align}
    P = \epsilon \cdot \widehat{P}, \label{eq:template}
\end{align}
where $\epsilon$ is a constant controlling the energy of the perturbation ($\epsilon=0.5$ in our experiments).

To mislead the P300 speller, one only needs to tamper with some specific signal periods according to the onset of the target stimuli. Because in a practical P300 speller the same row or column is never intensified successively, the perturbation template can last more than one intensification period. In our experiments, the template lasted $2\times175=350$ ms, i.e., two intensification periods.

\begin{figure*}[ht] \centering
    \includegraphics[width=\linewidth,clip]{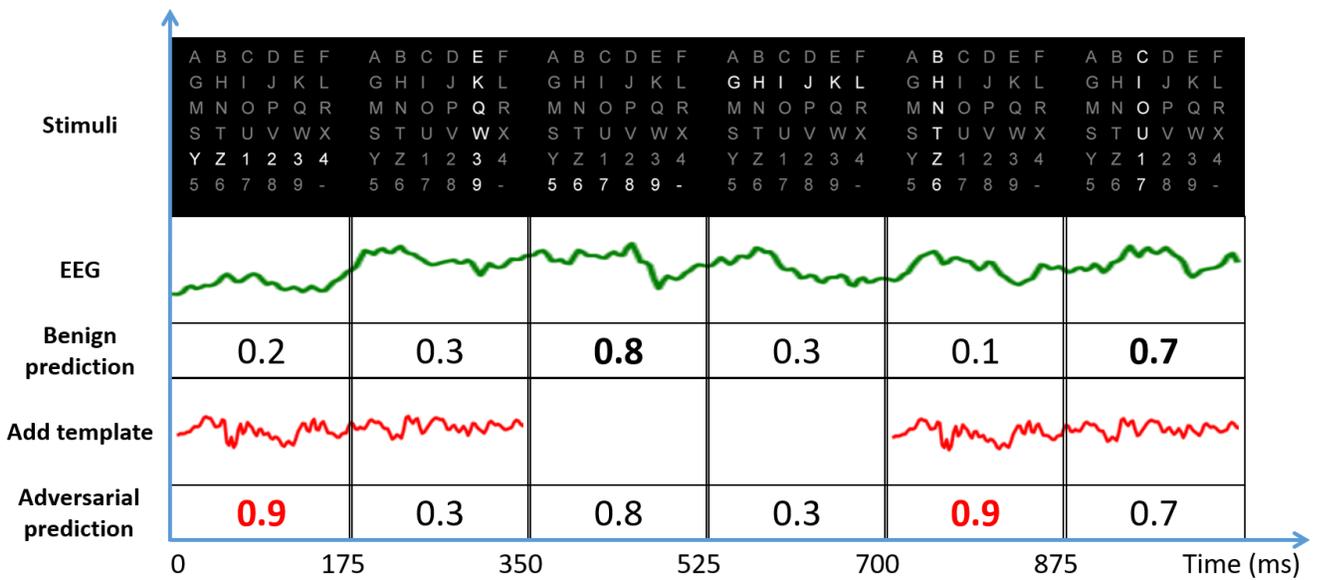}
\caption{Illustration of the attack procedure in the P300 protocol. The attacker character is \textit{Z}, whereas the user character is \textit{7}. For the benign EEG trial, the P300 speller can correctly identify that P300 is elicited by the intensifications of the last row and the third column. To mislead the P300 speller, adversarial perturbation template is added during the periods of 0-350ms and 700-1050ms, so that the fifth row and the second column are believed to elicit P300 with the highest probability. The added adversarial perturbation templates do not influence the results of the second and the last stimuli, because their corresponding periods are out of synchronization with the templates. As a result, the P300 speller misclassifies the perturbed trial to attacker character \textit{Z}. } \label{fig:Attack}
\end{figure*}

Figure~\ref{fig:Attack} illustrates the attack procedure. The benign EEG trial would output character \textit{7}, since the last row and the third column of the character matrix have the highest P300 probability, and their intersection is \textit{7}. However, after applying the perturbation template, the trial outputs the character \textit{Z}, because the fifth row and the second column have the highest P300 probability. Interestingly, the adversarial template acts like random noise when it is not synchronized with an intensification onset. As shown in Figure~\ref{fig:Attack}, the last 175 ms of the template does not influence the classification of the corresponding intensification.

\subsection{Attack the SSVEP Speller} \label{sect:MethodSSVEP}

There are two difficulties in attacking the victim model of the SSVEP speller. First, the victim model is not fixed, as the parameters of CCA vary in different EEG trials. Second, unlike the P300 speller whose base victim model only needs to classify the input into two classes, there are many more classes in the SSVEP speller. These make adversarial attacks of the SSVEP speller much more challenging.

The remedy was to generate the adversarial perturbation template $\delta_{\widehat{f}}\in\mathbb{R}^{N_e \times N_s}$, which can lead the SSVEP speller to output the attacker character of stimulation frequency $\widehat{f}$. For each user, we used the first block $\mathcal{D}=\{X_i\}_{i=1}^{N}$ to craft $\delta_{\widehat{f}}$, and the remaining five blocks to evaluate its attack performance.

According to the victim model, $\delta_{\widehat{f}}$ should be able to maximize $\rho(X+\delta_{\widehat{f}},Y_{\widehat{f}})$ in equation (\ref{eq:rho}), such that
\begin{align}
\arg\max_{f\in F} \rho(X+\delta_{\widehat{f}},Y_f) = \widehat{f}.
\end{align}
In other words, $\delta_{\widehat{f}}$ can be crafted by solving
\begin{align}
\max_{\delta_{\widehat{f}}} \ \sum_{X\in \mathcal{D}}\lambda_{\max}(S(X+\delta_{\widehat{f}},Y_{\widehat{f}})), \label{eq:delta1}
\end{align}
where $S(X,Y)$ is defined in equation~(\ref{eq:Sxy}).

Since $S(X+\delta_{\widehat{f}},Y_f)$ is not symmetric, it is difficult to calculate the derivatives of its largest eigenvalue, resulting in challenges in optimization. Because of the fact that the largest eigenvalue is always no smaller than the average of all eigenvalues:
\begin{align}
\lambda_{\max}(S(X+\delta_{\widehat{f}},Y_f)) &\geq \frac{1}{N_e}\sum\nolimits_j \lambda_{j}(S(X+\delta_{\widehat{f}},Y_f))\nonumber\\
& = \frac{1}{N_e}\mbox{tr}(S(X+\delta_{\widehat{f}},Y_f)),
\end{align}
instead of solving equation (\ref{eq:delta1}) directly, we can maximize its lower bound to reduce the optimization difficulty:
\begin{align}
\max_{\delta_{\widehat{f}}} \ \sum_{X\in \mathcal{D}} \mbox{tr}(S(X+\delta_{\widehat{f}}, Y_{\widehat{f}})). \label{eq:delta2}
\end{align}

Because the effective frequency band of SSVEP signals is 7-90~Hz, we introduced a new variable $\mathbf{r}_{\widehat{f}}$ so that
\begin{align}
\delta_{\widehat{f}}=\mbox{filt}(\mathbf{r}_{\widehat{f}}),
\end{align}
where $\mbox{filt}(\cdot)$ means retaining only the 7-90~Hz effective signal frequency components. As a result, we can ensure the integrity of the adversarial template during signal filtering. In addition, we added $\alpha \cdot \|\delta_{\widehat{f}}\|_F$ to penalize the energy of the perturbation, where $\alpha$ is the penalty coefficient.

Finally, the problem becomes:
\begin{align}
\min_{\mathbf{r}_{\widehat{f}}} \ -\sum_{X\in \mathcal{D}} \mbox{tr}(S(X+\mbox{filt}(\mathbf{r}_{\widehat{f}}), Y_{\widehat{f}}))+\alpha \cdot \|\mbox{filt}(\mathbf{r}_{\widehat{f}})\|_F. \label{eq:finaldelta}
\end{align}
Gradient descent was used to update $\mathbf{r}_{\widehat{f}}$, and the iteration stopped when the SPR was lower than a threshold, which was set to 25dB in our experiments.

\section*{DATA AVAILABILITY STATEMENT}
Publicly available BCI datasets were used in this study. The P300 speller dataset can be downloaded from \url{http://www.bbci.de/competition/iii/\#data_set_ii} (Dataset II). The P300 speller dataset of ALS patients was first used in \cite{Riccio2013} and can be downloaded from \url{http://bnci-horizon-2020.eu/database/data-sets} (P300 speller with ALS patients (008-2014)). The SSVEP dataset can be downloaded from \url{http://bci.med.tsinghua.edu.cn/download.html}. All source code is available on GitHub (\url{https://github.com/ZhangXiao96/Speller-Attacks}).

\clearpage

\begin{table*} \centering \setlength{\tabcolsep}{2.5mm}
    \caption{P300 speller attack results. Before attack: Baselines on clean EEG data (without adding any perturbations) and Gaussian-noise-perturbed EEG data, and the corresponding SPRs (dB). After attack: Average user/attacker scores/ITRs of the 36 attacker characters in target attacks, and the corresponding period and trial SPRs (dB).}   \label{tab:Tab1}
    \begin{tabular}{c|c|cc|ccc|cc|cc|cc}
        \toprule
        \multirow{3}{*}{Sub.} &Number    &\multicolumn{5}{c}{Before attack}   &\multicolumn{6}{|c}{After attack} \\ \cline{3-13}
                                 &of        &\multicolumn{2}{c}{Clean}  &\multicolumn{3}{|c}{Gaussian noise}    &\multicolumn{2}{|c}{User} &\multicolumn{2}{|c|}{Attacker} &Period & Trial\\ \cline{3-11}
                                 &repeats   &Score      &ITR            &Score      &ITR    &SPR    &Score      &ITR        &Score      &ITR    &   SPR & SPR        \\ \midrule
                                 &5         &0.64       &13.07          &0.65       &13.40  &20.8   &0.072      &0.248      &0.825      &19.8   &20.8 & 25.8  \\
        A                        &10        &0.85       &10.62          &0.84       &10.40  &21.0   &0.049      &0.052      &0.900      &11.7   &21.0 & 25.9  \\
                                 &15        &0.91       &\multicolumn{1}{r|}{8.03}          &0.92       &\multicolumn{1}{r}{8.19}  &21.0   &0.040      &0.021      &0.950      &\multicolumn{1}{r|}{8.7}   &21.0 & 25.9  \\ \midrule
                                 &5         &0.79       &18.41          &0.79       &18.41  &25.2   &0.107      &0.578      &0.713      &15.6   &25.2 & 30.2  \\
        B                        &10        &0.91       &11.96          &0.89       &11.50  &25.5   &0.061      &0.093      &0.860      &10.9   &25.5 & 30.4  \\
                                 &15        &0.93       &\multicolumn{1}{r|}{8.35}          &0.91       &\multicolumn{1}{r}{8.03}  &25.6   &0.049      &0.034      &0.907      &\multicolumn{1}{r|}{8.0}   &25.6 & 30.5  \\
        \bottomrule
    \end{tabular}
\end{table*}

\begin{table*} \centering \setlength{\tabcolsep}{2mm}
    \caption{SSVEP speller attack results. Before attack: Baselines on clean data (without adding any perturbations), Gaussian-noise-perturbed EEG data and periodic-noise-perturbed EEG data. After attack: Average user/attacker scores/ITRs of 40 attacker characters in target attacks, and the corresponding SPRs (dB).}   \label{tab:Tab2}
    \begin{tabular}{c|cc|cc|cc|c|cr|cc|c}
        \toprule
        \multirow{3}{*}{Sub.}    &\multicolumn{7}{c|}{Before attack}    &\multicolumn{5}{c}{After attack}\\ \cline{2-13}
                &\multicolumn{2}{c}{Clean}  &\multicolumn{2}{|c}{Gaussian Noise} &\multicolumn{2}{|c|}{ S/C Periodic Noise} &\multirow{2}{*}{SPR} &\multicolumn{2}{c}{User}  &\multicolumn{2}{|c|}{Attacker} &\multirow{2}{*}{SPR}\\ \cline{2-7} \cline{9-12}
                &Score      &ITR            &Score      &ITR       &Score      &ITR     &       &Score      &\multicolumn{1}{c|}{ITR}        &Score      &ITR       &    \\ \midrule
        3  &0.88  &182.5  &0.88  &181.6  &0.71/0.87  &129.0/178.6  &25.0  &0.44  &61.1  &0.58  &\multicolumn{1}{r|}{93.3}     &25.3\\
        4  &0.90  &186.9  &0.90  &187.1  &0.68/0.87  &121.0/177.5  &25.0  &0.07  &2.3  &0.95  &210.1  &25.7\\
        12  &0.90  &188.8  &0.90  &188.0  &0.78/0.86  &150.0/174.3  &25.0  &0.26  &26.6  &0.75  &139.5  &25.5\\
        22  &0.82  &160.0  &0.79  &150.2  &0.74/0.75  &137.5/140.7  &25.0  &0.11  &6.1  &0.91  &191.0  &25.1\\
        25  &0.90  &189.1  &0.89  &184.1  &0.84/0.87  &168.3/177.2  &25.0  &0.78  &148.2  &0.17  &\multicolumn{1}{r|}{13.8}  &26.7\\
        26  &0.90  &187.8  &0.88  &180.3  &0.58/0.84  &\multicolumn{1}{r|}{94.4/168.1}  &25.0  &0.03  &0.1  &1.00  &229.9  &24.8\\
        32  &0.87  &176.9  &0.87  &179.6  &0.59/0.82  &\multicolumn{1}{r|}{97.2/163.6}  &25.0  &0.03  &0.0  &1.00  &231.4  &24.9\\
        34  &0.80  &154.7  &0.79  &151.8  &0.48/0.72  &\multicolumn{1}{r|}{66.8/130.4}  &25.0  &0.03  &0.0  &1.00  &231.2  &25.9\\
        \bottomrule
    \end{tabular}
\end{table*}

\clearpage
\textbf{\Large Supplementary Information}\\
\section{The victim model of the P300 speller} \label{sect:P300model}

The details of the victim model of the P300 speller are introduced.

\subsection{xDAWN spatial filters}

The original xDAWN filter \cite{Rivet2009} was designed for P300 evoked potentials by enhancing the target response with respect to the non-target response. We used a generalized version in our experiments, which was implemented in \emph{pyRiemann}\footnote{https://pyriemann.readthedocs.io/en/latest/index.html}.

More specifically, let $\mathbf{D}=\{(X_i, y_i)\}_{i=1}^N$ be the training set, where $X_i \in \mathbb{R}^{N_e \times N_s}$ is the $i$-th mean-centered EEG epoch ($N_e$ is the number of channels, and $N_s$ the number of time domain samples), and $y_i \in \{0, 1\}$ its corresponding label ($0$ for \emph{non-target}, and $1$ for \emph{target}). The average epoch $\overline{X}_c$, $c \in \{0, 1\}$, is first calculated. Spatial filters $U_c \in \mathbb{R}^{N_f \times N_e}$ were then designed to maximize the signal to signal-plus-noise ratio for each class:
\begin{align}
U_c = \mathop{\mbox{arg max}}\limits_{U}\frac{\mbox{tr}\left(U\overline{X}_c\overline{X}_c^TU^T\right)}{\mbox{tr}\left(UX_{all}X_{all}^TU^T\right)}, \label{eq:xdawn}
\end{align}
where $N_f$ is the number of filters ($N_f=8$ was used in our experiments), $X_{all}$ is obtained by concatenating all EEG epochs in $\mathbf{D}$ along the channels, and $tr$ is the trace of a matrix. Generalized eigenvalue decomposition can be used to solve equation (\ref{eq:xdawn}).

After obtaining the filters for both classes, the concatenated spatial filters $U = \left[U_0; U_1\right]$ can be used to filter each EEG epoch:
\begin{align}
\widetilde{X}_i = UX_i.
\end{align}

\subsection{Tangent space projection}

Covariance matrices of the EEG trials are widely-used in BCIs. However, they lie on a Riemannian manifold of Symmetric Positive Definite (SPD) matrices, and hence cannot be directly used by Euclidean space classifiers, such as Logistic Regression and Support Vector Machines. To solve this problem, the covariance matrices are projected onto the Euclidean tangent space of a reference SPD matrix, and then the vectorized features are used by Euclidean space classifiers.

More specifically, we first calculate the augmented covariance matrix $C_i$ for each $\widetilde{X}_i$:
\begin{align}
C_i = \left[\begin{array}{cc}
                         ZZ^T, &Z\widetilde{X}_i^T \\
                         \widetilde{X}_iZ^T, &\widetilde{X}_i\widetilde{X}_i^T
                       \end{array}\right],
\end{align}
where $Z=\left[U\overline{X}_0; U\overline{X}_1\right]$. Then, $C_i$ is projected onto the tangent space of the reference SPD matrix $C_f$, which is the geometric mean of $\{C_i\}_{i=1}^N$, i.e.,
\begin{align}
C_f = \arg \mathop{\min}\limits_{C}{\left(\sum_{i=1}^{N} \delta(C,C_i)^2\right)},
\end{align}
where $\delta(C_A, C_B)$ is the Affine Invariant Riemannian Metric distance:
\begin{align}
\delta(C_A,C_B)={\Big\|\mbox{logm}\left(C_A^{-1/2}C_BC_A^{-1/2}\right)\Big\|}.
\end{align}

The vectorized features are:
\begin{align}
\mathbf{s}_i=\mbox{upper}\left(\mbox{logm}\left(C_f^{-1/2}C_iC_f^{-1/2}\right)\right),
\end{align}
where $\mbox{upper}(\cdot)$ vectorizes the upper triangular part of a symmetric matrix. A weight of $\sqrt{2}$ is applied to the off-diagonal elements, and a weight of $1$ to the rest, during the vectorization.  $\mathbf{s}_i$ can then be fed into any Euclidean space classifier.

\section{Canonical correlation analysis (CCA)} \label{sect:CCA}

This section introduces CCA, which can be used to extract the underlying correlation between two time series.

\subsection{Problem setup}

Let $X\in\mathbb{R}^{C_1 \times N}$ and $Y\in\mathbb{R}^{C_2 \times N}$ be two multi-channel time series, where $C_1$ and $C_2$ represent the number of channels, and $N$ the number of time domain samples. $X$ and $Y$ are $z$-normalized in each channel.

The main idea of CCA is to find a pair of canonical variables, denoted as $\mathbf{a}\in\mathbb{R}^{C_1\times 1}$ and $\mathbf{b}\in\mathbb{R}^{C_2\times 1}$, for $X$ and $Y$ respectively, so that the correlation coefficient $\rho$ between $\mathbf{a}^TX$ and $\mathbf{b}^TY$ can be maximized. The problem can be mathematically formulated as:
\begin{align}
\max_{\mathbf{a},\mathbf{b}} \frac{\mathbf{a}^TXY^T\mathbf{b}}{\sqrt{\mathbf{a}^TXX^T\mathbf{a}}\sqrt{\mathbf{b}^TYY^T\mathbf{b}}},
\end{align}
which can be re-expressed as:
\begin{align}
\max_{\mathbf{a},\mathbf{b}}\quad&\mathbf{a}^TXY^T\mathbf{b},  \label{eq:CCA}\\
\mbox{s.t.}\quad&\mathbf{a}^TXX^T\mathbf{a}=1,\mathbf{b}^TYY^T\mathbf{b}=1. \nonumber
\end{align}

\subsection{Solution of CCA}

There are several approaches to solve equation (\ref{eq:CCA}). Here we introduce the Lagrange multiplier method \cite{Hestenes1969}.

Denote $AB^T$ by $S_{AB}$. Then, equation (\ref{eq:CCA}) can be rewritten as:
\begin{align}
\max_{\mathbf{a}, \mathbf{b}}\quad&\mathbf{a}^TS_{XY}\mathbf{b},  \label{eq:CCAS}\\
\mbox{s.t.}\quad&\mathbf{a}^TS_{XX}^T\mathbf{a}=1, \mathbf{b}^TS_{YY}^T\mathbf{b}=1. \nonumber
\end{align}

According to the Lagrange multiplier method, equation (\ref{eq:CCAS}) is equivalent to $\displaystyle\max_{\mathbf{a},\mathbf{b},\lambda,\theta}J(\mathbf{a},\mathbf{b},\lambda,\theta)$, where:
\begin{align}
J(\mathbf{a},\mathbf{b},\lambda,\theta)=\mathbf{a}^TS_{XY}\mathbf{b}-\frac{\lambda}{2}(\mathbf{a}^TS_{XX}\mathbf{a}-1)-\frac{\theta}{2}(\mathbf{b}^TS_{YY}\mathbf{b}-1).  \label{eq:CCAJ}
\end{align}

By setting the first partial derivatives to zero, i.e.,
\begin{align}
\nabla_{\mathbf{a}}J&=S_{XY}\mathbf{b}-\lambda\cdot S_{XX}\mathbf{a}=0, \label{eq:con1}\\
\nabla_{\mathbf{b}}J&=S_{YX}\mathbf{a}-\theta\cdot  S_{YY}\mathbf{b}=0, \label{eq:con2}\\
\frac{\partial J}{\partial \lambda}&=-\frac{1}{2}(\mathbf{a}^TS_{XX}\mathbf{a}-1)=0,  \\
\frac{\partial J}{\partial \theta}&=-\frac{1}{2}(\mathbf{a}^TS_{YY}\mathbf{b}-1)=0,
\end{align}
we have
\begin{align}
\lambda=\theta=\mathbf{a}^TS_{XY}\mathbf{b}.  \label{eq:lambda}
\end{align}
It should be noted that equation (\ref{eq:lambda}) is also the definition of the correlation coefficient $\rho$.

According to equations (\ref{eq:con1}) and (\ref{eq:con2}), we have:
\begin{align}
S_{XX}^{-1}S_{XY}S_{YY}^{-1}S_{YX}\mathbf{a}=\lambda^2\mathbf{a}=\rho^2\mathbf{a},
\end{align}
which implies that $\rho^2$ equals the largest eigenvalue of $S_{XX}^{-1}S_{XY}S_{YY}^{-1}S_{YX}$, and $\mathbf{a}$ is the corresponding eigenvector.

$\mathbf{b}$ can be obtained in a similar way.

\section{Security of a P300 speller for Amyotrophic Lateral Sclerosis (ALS) patients}

We performed additional experiments to investigate how adversarial perturbations impact ALS patients on P300 Spellers \cite{Riccio2013}. The eight-channel (Fz, Cz, Pz, Oz, P3, P4, PO7, and PO8) EEG signals were recorded from eight ALS patients. The EEG data were digitized at 256 Hz, bandpass filtered to 0.1-30 Hz, and then $z$-normalized for each channel. For each subject, there were 21 characters for training and 14 for testing. Each character corresponds to a set of 12 random intensifications, which were repeated 20 times. Each intensification lasted for 125 ms, followed by a 125 ms blank. In our experiments, 10 repeats were utilized to output a character during the test.

We applied the same Riemannian geometry based approach to recognizing the existence of P300 potentials. The only difference from the previous study was that the number of xDAWN spatial filters was eight. As shown in the `Before attack' panel of Table~\ref{tab:ALSResult}, the victim models demonstrated good performance without attacks, and also high robustness to Gaussian noise perturbations. However, the `After attack' panel shows that all user scores and ITRs were more or less reduced after adversarial perturbations. For half of the subjects (subjects 1, 2, 5 and 7), the user scores and ITRs approached zero, i.e., the P300 speller became almost completely useless, indicating a serious security concern of P300 spellers to ALS patients.

\begin{table*}[h] \centering \setlength{\tabcolsep}{2.5mm}
    \caption{P300 speller attack results for eight ALS patients. Before attack: Baselines on clean EEG data (without adding any perturbations) and Gaussian-noise-perturbed EEG data, and the corresponding SPRs (dB). After attack: Average user/attacker scores/ITRs of the 36 attacker characters in target attacks, and the corresponding period and trial SPRs (dB). $\epsilon=0.8$ for all the perturbations.}   \label{tab:ALSResult}
    \begin{tabular}{c|cc|ccc|cc|cc|cc}
        \toprule
        \multirow{3}{*}{Sub.} &\multicolumn{5}{c}{Before attack}   &\multicolumn{6}{|c}{After attack} \\ \cline{2-12}
                                 &\multicolumn{2}{c}{Clean}  &\multicolumn{3}{|c}{Gaussian noise}    &\multicolumn{2}{|c}{User} &\multicolumn{2}{|c|}{Attacker} &Period & Trial\\ \cline{2-10}
                                 &Score      &ITR      &Score   &ITR    &SPR    &Score      &ITR    &Score  &ITR    &SPR & SPR \\ \midrule
        1  &0.79  &\multicolumn{1}{r|}{6.57}   &0.79  &\multicolumn{1}{r}{6.57}  &22.6  &0.03  &0.04  &1.00  &10.22  &22.6 &27.4\\
        2  &0.93  & \multicolumn{1}{r|}{8.76}   &0.93  &\multicolumn{1}{r}{8.76}  &22.4  &0.10  &0.26  &0.74  &\multicolumn{1}{r|}{6.09}   &22.4 &27.5\\
        3  &1.00  &10.22  &1.00  &10.22 &22.9  &0.53  &3.59  &0.17  &\multicolumn{1}{r|}{0.67}   &22.9 &27.7\\
        4  &1.00  &10.22  &0.93  &\multicolumn{1}{r}{8.76}  &23.1  &0.45  &2.83  &0.22  &\multicolumn{1}{r|}{1.10}   &23.1 &27.9\\
        5  &1.00  &10.22  &1.00  &10.22 &22.2  &0.05  &0.12  &0.86  &\multicolumn{1}{r|}{7.72}   &22.2 &27.1\\
        6  &0.93  &\multicolumn{1}{r|}{8.76}   &0.86  &\multicolumn{1}{r}{7.60}  &22.4  &0.21  &0.86  &0.44  &\multicolumn{1}{r|}{2.75}   &22.4 &27.2\\
        7  &1.00  &10.22   &1.00  &10.22 &22.9  &0.03  &0.05  &0.96  &\multicolumn{1}{r|}{9.45}   &22.9 &27.7\\
        8  &1.00  &10.22   &1.00  &10.22 &23.1  &0.98  &9.73  &0.03  &\multicolumn{1}{r|}{0.05}   &23.1 &28.0\\
        \bottomrule
    \end{tabular}
\end{table*}

\section{Transferability of adversarial perturbations}

We have mentioned that a limitation of the attack approaches is that they require some subject-/model-specific information to construct adversarial perturbation templates. One possible solution to alleviate this problem is to enhance the transferability of adversarial perturbations: the attacker can generate adversarial perturbations based on EEG signals gathered by himself/herself, or any model he/she chooses to use, and then utilize them to attack another subject/model. Here, we present some experimental results on the cross-subject and cross-model transferability of our adversarial perturbations (for P300 spellers, the ALS patient dataset was used due to its large number of subjects). We found that adversarial perturbations for SSVEP spellers seem to have better transferability than P300 spellers.

\subsection{Cross-subject transferability}

\begin{figure}[h]\centering
\includegraphics[width=\linewidth,clip]{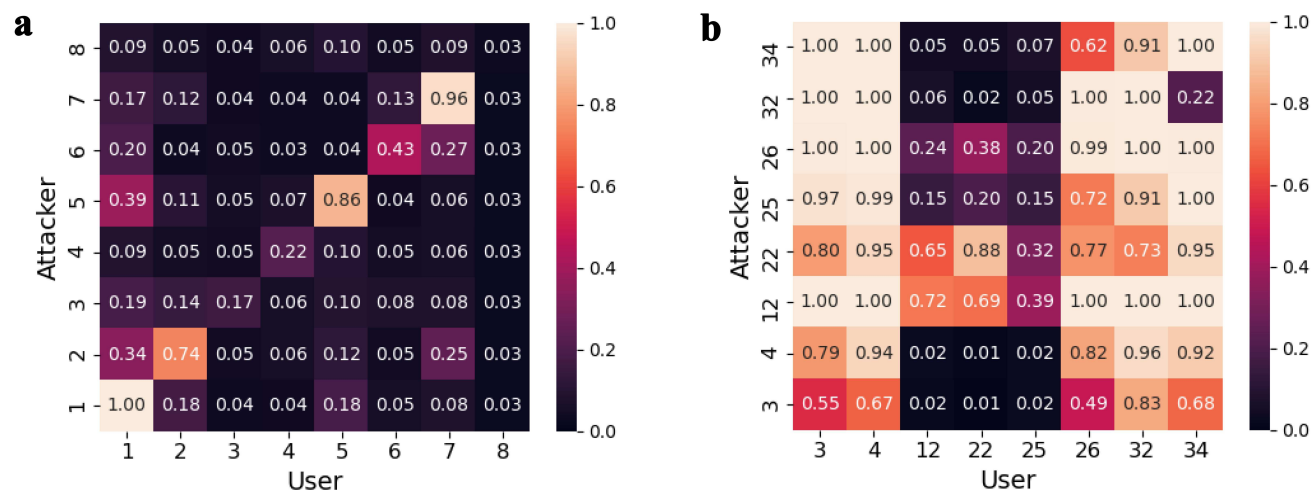} \\
\caption{Cross-subject transferability of adversarial perturbations. The heatmap shows the average attacker scores when using the adversarial perturbations of one subject to attack another subject. \textbf{a}, attacker scores for the P300 speller. \textbf{b}, attacker scores for the SSVEP speller.} \label{fig:UserTransfer}
\end{figure}

We used adversarial perturbations generated from one subject to attack the victim model of another subject.
Figure~\ref{fig:UserTransfer} shows the average attacker scores of cross-subject attacks. There was almost no cross-subject transferability of adversarial perturbation templates for the P300 spellers, whereas perturbations for SSVEP spellers can usually successfully attack the victim models of different subjects. Additionally, some subjects were much more robust to transfer attacks, e.g., Subjects 12, 22 and 25 in Figure~\ref{fig:UserTransfer}b.

Why adversarial perturbation templates demonstrated poor cross-subject transferability for the P300 spellers would be investigated in more depth in our future research.

\subsection{Cross-model transferability}

Cross-model transferability requires adversarial perturbations to be able to attack different EEG classification pipelines, which means the attacker does not need access to victim models anymore, implying a more serious threat to the security of BCI spellers. This subsection presents the attack performance of our generated adversarial perturbations on new EEG classification pipelines.

For the P300 spellers, the new classification pipeline consisted of xDAWN filtering and Logistic Regression classification, and the adversarial perturbation templates were again generated from the Riemannian geometry based approach. The `Before attack' panel of Table~\ref{tab:P300Cross} shows that the new pipeline had high classification accuracy without attacks, and it was also robust to Gaussian noise. The `After attack' panel shows that the new pipeline can still be manipulated by adversarial perturbation templates constructed from a different pipeline, though not as much as that in Table~\ref{tab:ALSResult}. Comparing the attack performances in Tables~\ref{tab:ALSResult} and \ref{tab:P300Cross}, it seems that an adversarial perturbation template with better attack performance on the model it was generated from may also have better cross-model transferability to attack another model.

\begin{table*}[ht] \centering \setlength{\tabcolsep}{2.5mm}
    \caption{P300 speller cross-model attack results for eight ALS patients. The victim model (xDAWN and Logistic Regression) was different from the attacker model (a Riemannian geometry based approach), based on which adversarial perturbations were generated. Before attack: Baselines on clean EEG data (without adding any perturbations) and Gaussian-noise-perturbed EEG data, and the corresponding SPRs (dB). After attack: Average user/attacker scores/ITRs of the 36 attacker characters in target attacks, and the corresponding period and trial SPRs (dB). $\epsilon=0.8$ for all perturbations.}   \label{tab:P300Cross}
    \begin{tabular}{c|cc|ccc|cc|cc|cc}
        \toprule
        \multirow{3}{*}{Sub.} &\multicolumn{5}{c}{Before attack}   &\multicolumn{6}{|c}{After attack} \\ \cline{2-12}
        &\multicolumn{2}{c}{Clean}  &\multicolumn{3}{|c}{Gaussian noise}    &\multicolumn{2}{|c}{User} &\multicolumn{2}{|c|}{Attacker} &Period & Trial\\ \cline{2-10}
        &Score      &ITR      &Score   &ITR    &SPR    &Score      &ITR    &Score  &ITR    &SPR & SPR \\ \midrule
        1  &0.86  &\multicolumn{1}{r|}{7.60}   &0.86  &\multicolumn{1}{r}{7.60}  &22.6  &0.03  &0.04  &1.00  &10.22  &22.6 &27.4\\
        2  &1.00  &10.22   &1.00  &10.22  &22.4  &0.53  &3.56  &0.24  &\multicolumn{1}{r|}{0.99}   &22.4 &27.5\\
        3  &1.00  &10.22  &1.00  &10.22 &22.9  &0.62  &4.66  &0.15  &\multicolumn{1}{r|}{0.60}   &22.9 &27.7\\
        4  &0.79  &\multicolumn{1}{r|}{6.57}  &0.79  &\multicolumn{1}{r}{6.57}  &23.1  &0.49  &3.21  &0.20  &\multicolumn{1}{r|}{0.91}   &23.1 &27.9\\
        5  &0.86  &\multicolumn{1}{r|}{7.60}  &0.86  &\multicolumn{1}{r}{7.60} &22.2  &0.17  &0.60  &0.60  &\multicolumn{1}{r|}{4.31}   &22.2 &27.1\\
        6  &1.00  &10.22   &1.00  &10.22  &22.4  &0.36  &1.90  &0.31  &\multicolumn{1}{r|}{1.66}   &22.4 &27.2\\
        7  &1.00  &10.22   &1.00  &10.22 &22.9  &0.17  &0.61  &0.53  &\multicolumn{1}{r|}{3.59}   &22.9 &27.7\\
        8  &1.00  &10.22   &1.00  &10.22 &23.1  &0.94  &9.04  &0.04  &\multicolumn{1}{r|}{0.06}   &23.1 &28.0\\
        \bottomrule
    \end{tabular}
\end{table*}

For the SSVEP spellers, we utilized Filter Bank Canonical Correlation Analysis (FBCCA)\footnote{Our implementation was adapted from \url{https://github.com/hisunjiang/CCAforSSVEP}.} as our new victim model \cite{Chen2015}. Table~\ref{tab:SSVEPCross} shows the baseline performance of FBCCA and the attack performance of adversarial perturbations (generated from CCA) on this model. FBCCA demonstrated promising performance on clean and randomly perturbed EEG signals. However, adversarial perturbations generated from CCA can still manipulate the output characters of FBCCA, verifying that cross-model transferability also exists in the SSVEP spellers.

\begin{table*}[h] \centering \setlength{\tabcolsep}{2mm}
\caption{SSVEP speller cross-model attack results. The victim model (FBCCA) was different from the attacker model (CCA), based on which adversarial perturbations were generated. Before attack: Baselines on clean data (without adding any perturbations), Gaussian-noise-perturbed EEG data and periodic-noise-perturbed EEG data (single/compound). After attack: Average user/attacker scores/ITRs of 40 attacker characters in target attacks, and the corresponding SPRs (dB).}   \label{tab:SSVEPCross}
\begin{tabular}{c|cc|cc|cc|c|cr|cc|c}
        \toprule
        \multirow{3}{*}{Sub.}    &\multicolumn{7}{c|}{Before attack}    &\multicolumn{5}{c}{After attack}\\ \cline{2-13}
        &\multicolumn{2}{c}{Clean}  &\multicolumn{2}{|c}{Gaussian Noise} &\multicolumn{2}{|c|}{ S/C Periodic Noise} &\multirow{2}{*}{SPR} &\multicolumn{2}{c}{User}  &\multicolumn{2}{|c|}{Attacker} &\multirow{2}{*}{SPR}\\ \cline{2-7} \cline{9-12}
        &Score      &ITR            &Score      &ITR       &Score      &ITR     &       &Score      &\multicolumn{1}{c|}{ITR}        &Score      &ITR       &    \\ \midrule
        3  &0.98  &218.7  &0.98  &219.0  &0.96/0.97  &212.7/217.7  &25.0  &0.06  &2.3  &0.92  &204.2     &25.3\\
        4  &0.88  &181.4  &0.88  &182.2  &0.84/0.88  &169.0/180.1  &25.0  &0.03  &0.0  &1.00  &230.6  &25.7\\
        12  &0.85  &173.3  &0.82  &163.1  &0.81/0.81  &159.9/159.9  &25.0  &0.05  &2.2  &0.97  &219.0  &25.5\\
        22  &0.95  &207.4  &0.95  &208.1  &0.94/0.95  &204.7/206.4  &25.0  &0.03  &0.4  &0.99  &228.6  &25.1\\
        25  &0.87  &176.8  &0.84  &168.1  &0.83/0.84  &163.4/166.4  &25.0  &0.10  &7.8  &0.90  &194.0  &26.7\\
        26  &0.87  &177.9  &0.86  &174.1  &0.75/0.82  &139.8/159.6  &25.0  &0.03  &0.0  &1.00  &231.4  &24.8\\
        32  &0.88  &181.2  &0.87  &180.1  &0.80/0.85  &157.1/170.5  &25.0  &0.03  &0.0  &1.00  &231.4  &24.9\\
        34  &0.88  &180.7  &0.85  &171.3  &0.72/0.80  &132.6/155.9  &25.0  &0.03  &0.0  &0.98  &226.4  &25.9\\
        \bottomrule
    \end{tabular}
\end{table*}

The key to the transferability is to find the most common patterns shared by different models, hence the adversarial perturbations affecting these patterns can attack as many models as possible. From this point of view, generating adversarial perturbations based on an ensemble of multiple models presents a serious threat to the security of BCIs. We will explore this in our future research.

\section{Additional Figures}

\begin{figure}[h]\centering
    \includegraphics[width=\linewidth,clip]{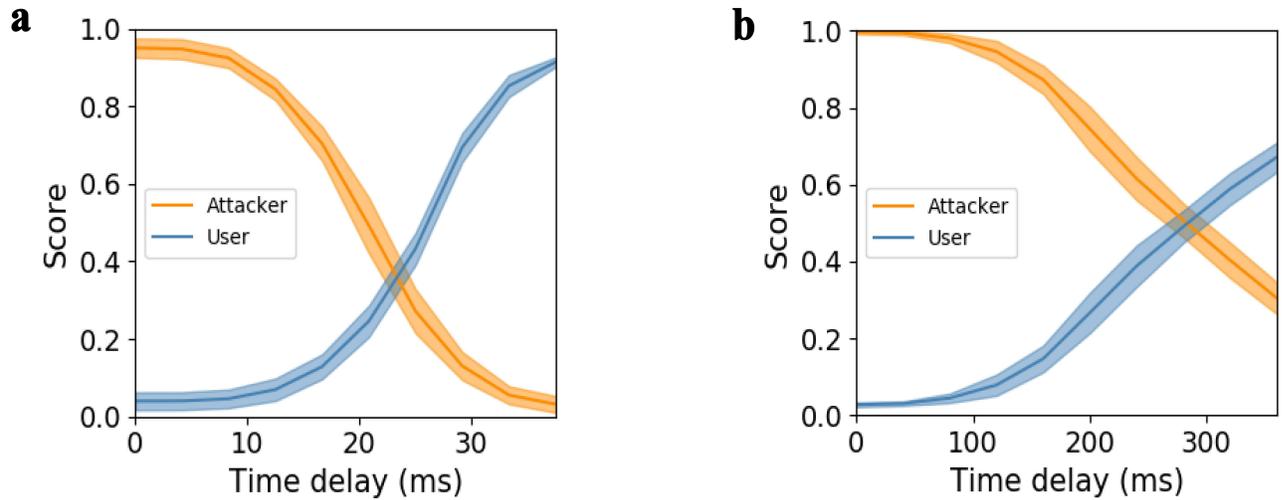} \\
\caption{User and attacker scores with respect to the synchronization time delay. The curve represents the mean of all attacker characters, and the shadow the standard deviation. \textbf{a}, scores for the P300 speller, where 100 test trials for Subject~A were perturbed to be misclassified as each of the 36 attacker characters. \textbf{b}, scores for the SSVEP speller, where $5\times40=200$ test trials for Subject~26 were perturbed to be misclassified as each of the 40 attacker characters.} \label{fig:TimeDelay}
\end{figure}

Figure~\ref{fig:TimeDelay} shows how the synchronization time delay affects the attack performance.

\begin{thebibliography}{10}
\expandafter\ifx\csname url\endcsname\relax
  \def\url#1{\texttt{#1}}\fi
\expandafter\ifx\csname urlprefix\endcsname\relax\def\urlprefix{URL }\fi
\providecommand{\bibinfo}[2]{#2}
\providecommand{\eprint}[2][]{\url{#2}}

\bibitem{Graimann2009}
\bibinfo{author}{Graimann, B.}, \bibinfo{author}{Allison, B.} \&
  \bibinfo{author}{Pfurtscheller, G.}
\newblock \emph{\bibinfo{title}{Brain-computer interfaces: {A} gentle
  introduction}}, \bibinfo{pages}{1--27} (\bibinfo{publisher}{Springer},
  \bibinfo{address}{Berlin, Heidelberg}, \bibinfo{year}{2009}).

\bibitem{Lin2017}
\bibinfo{author}{Lin, C.~T.} \emph{et~al.}
\newblock \bibinfo{title}{{EEG}-based brain-computer interfaces: {A} novel
  neurotechnology and computational intelligence method}.
\newblock \emph{\bibinfo{journal}{IEEE Systems, Man, and Cybernetics Magazine}}
  \textbf{\bibinfo{volume}{3}}, \bibinfo{pages}{16--26} (\bibinfo{year}{2017}).

\bibitem{drwuEA2020}
\bibinfo{author}{He, H.} \& \bibinfo{author}{Wu, D.}
\newblock \bibinfo{title}{Transfer learning for brain-computer interfaces: A
  {Euclidean} space data alignment approach}.
\newblock \emph{\bibinfo{journal}{{IEEE} Trans. on Biomedical Engineering}}
  \textbf{\bibinfo{volume}{67}}, \bibinfo{pages}{399--410}
  (\bibinfo{year}{2020}).

\bibitem{Chavarriaga2010}
\bibinfo{author}{Chavarriaga, R.} \& \bibinfo{author}{Mill{\'a}n, J. d.~R.}
\newblock \bibinfo{title}{Learning from {EEG} error-related potentials in
  noninvasive brain-computer interfaces}.
\newblock \emph{\bibinfo{journal}{IEEE Trans. on Neural Systems and
  Rehabilitation Engineering}} \textbf{\bibinfo{volume}{18}},
  \bibinfo{pages}{381--388} (\bibinfo{year}{2010}).

\bibitem{Nicolas-Alonso2012}
\bibinfo{author}{Nicolas-Alonso, L.~F.} \& \bibinfo{author}{Gomez-Gil, J.}
\newblock \bibinfo{title}{Brain computer interfaces, a review}.
\newblock \emph{\bibinfo{journal}{Sensors}} \textbf{\bibinfo{volume}{12}},
  \bibinfo{pages}{1211--1279} (\bibinfo{year}{2012}).

\bibitem{Farwell1988}
\bibinfo{author}{Farwell, L.~A.} \& \bibinfo{author}{Donchin, E.}
\newblock \bibinfo{title}{Talking off the top of your head: {T}oward a mental
  prosthesis utilizing event-related brain potentials}.
\newblock \emph{\bibinfo{journal}{Electroencephalography and Clinical
  Neurophysiology}} \textbf{\bibinfo{volume}{70}}, \bibinfo{pages}{510--523}
  (\bibinfo{year}{1988}).

\bibitem{Chen2015a}
\bibinfo{author}{Chen, X.} \emph{et~al.}
\newblock \bibinfo{title}{High-speed spelling with a noninvasive brain-computer
  interface}.
\newblock \emph{\bibinfo{journal}{Proc. National Academy of Sciences}}
  \textbf{\bibinfo{volume}{112}}, \bibinfo{pages}{E6058--E6067}
  (\bibinfo{year}{2015}).

\bibitem{Sutton1965}
\bibinfo{author}{Sutton, S.}, \bibinfo{author}{Braren, M.},
  \bibinfo{author}{Zubin, J.} \& \bibinfo{author}{John, E.~R.}
\newblock \bibinfo{title}{Evoked-potential correlates of stimulus uncertainty}.
\newblock \emph{\bibinfo{journal}{Science}} \textbf{\bibinfo{volume}{150}},
  \bibinfo{pages}{1187--1188} (\bibinfo{year}{1965}).

\bibitem{Donchin2000}
\bibinfo{author}{Donchin, E.}, \bibinfo{author}{Spencer, K.~M.} \&
  \bibinfo{author}{Wijesinghe, R.}
\newblock \bibinfo{title}{The mental prosthesis: {A}ssessing the speed of a
  {P}300-based brain-computer interface}.
\newblock \emph{\bibinfo{journal}{IEEE Trans. on Rehabilitation Engineering}}
  \textbf{\bibinfo{volume}{8}}, \bibinfo{pages}{174--179}
  (\bibinfo{year}{2000}).

\bibitem{Meinicke2003}
\bibinfo{author}{Meinicke, P.}, \bibinfo{author}{Kaper, M.},
  \bibinfo{author}{Hoppe, F.}, \bibinfo{author}{Heumann, M.} \&
  \bibinfo{author}{Ritter, H.}
\newblock \bibinfo{title}{Improving transfer rates in brain computer
  interfacing: {A} case study}.
\newblock In \emph{\bibinfo{booktitle}{Proc. Advances in Neural Information
  Processing Systems}}, \bibinfo{pages}{1131--1138} (\bibinfo{address}{BC,
  Canada}, \bibinfo{year}{2003}).

\bibitem{Xu2004}
\bibinfo{author}{Xu, N.} \emph{et~al.}
\newblock \bibinfo{title}{{BCI} competition 2003-data set {II}b: {E}nhancing
  {P}300 wave detection using {ICA}-based subspace projections for {BCI}
  applications}.
\newblock \emph{\bibinfo{journal}{IEEE Trans. on Biomedical Engineering}}
  \textbf{\bibinfo{volume}{51}}, \bibinfo{pages}{1067--1072}
  (\bibinfo{year}{2004}).

\bibitem{Guan2004}
\bibinfo{author}{Guan, C.}, \bibinfo{author}{Thulasidas, M.} \&
  \bibinfo{author}{Wu, J.}
\newblock \bibinfo{title}{High performance {P}300 speller for brain-computer
  interface}.
\newblock In \emph{\bibinfo{booktitle}{Proc. IEEE Int'l Workshop on Biomedical
  Circuits and Systems}}, \bibinfo{pages}{S3--5} (\bibinfo{address}{Singapore},
  \bibinfo{year}{2004}).

\bibitem{Polich2007}
\bibinfo{author}{Polich, J.}
\newblock \bibinfo{title}{Updating {P}300: {A}n integrative theory of {P}3a and
  {P}3b}.
\newblock \emph{\bibinfo{journal}{Clinical Neurophysiology}}
  \textbf{\bibinfo{volume}{118}}, \bibinfo{pages}{2128--2148}
  (\bibinfo{year}{2007}).

\bibitem{Chapman1964}
\bibinfo{author}{Chapman, R.~M.} \& \bibinfo{author}{Bragdon, H.~R.}
\newblock \bibinfo{title}{Evoked responses to numerical and non-numerical
  visual stimuli while problem solving}.
\newblock \emph{\bibinfo{journal}{Nature}} \textbf{\bibinfo{volume}{203}},
  \bibinfo{pages}{1155} (\bibinfo{year}{1964}).

\bibitem{Sutton1967}
\bibinfo{author}{Sutton, S.}, \bibinfo{author}{Tueting, P.},
  \bibinfo{author}{Zubin, J.} \& \bibinfo{author}{John, E.~R.}
\newblock \bibinfo{title}{Information delivery and the sensory evoked
  potential}.
\newblock \emph{\bibinfo{journal}{Science}} \textbf{\bibinfo{volume}{155}},
  \bibinfo{pages}{1436--1439} (\bibinfo{year}{1967}).

\bibitem{Beverina2003}
\bibinfo{author}{Beverina, F.} \emph{et~al.}
\newblock \bibinfo{title}{User adaptive {BCI}s: {SSVEP} and {P}300 based
  interfaces.}
\newblock \emph{\bibinfo{journal}{PsychNology Journal}}
  \textbf{\bibinfo{volume}{1}}, \bibinfo{pages}{331--354}
  (\bibinfo{year}{2003}).

\bibitem{Wang2008}
\bibinfo{author}{Wang, Y.}, \bibinfo{author}{Gao, X.}, \bibinfo{author}{Hong,
  B.}, \bibinfo{author}{Jia, C.} \& \bibinfo{author}{Gao, S.}
\newblock \bibinfo{title}{Brain-computer interfaces based on visual evoked
  potentials}.
\newblock \emph{\bibinfo{journal}{IEEE Engineering in Medicine and Biology
  Magazine}} \textbf{\bibinfo{volume}{27}}, \bibinfo{pages}{64--71}
  (\bibinfo{year}{2008}).

\bibitem{Vialatte2010}
\bibinfo{author}{Vialatte, F.-B.}, \bibinfo{author}{Maurice, M.},
  \bibinfo{author}{Dauwels, J.} \& \bibinfo{author}{Cichocki, A.}
\newblock \bibinfo{title}{Steady-state visually evoked potentials: {F}ocus on
  essential paradigms and future perspectives}.
\newblock \emph{\bibinfo{journal}{Progress in Neurobiology}}
  \textbf{\bibinfo{volume}{90}}, \bibinfo{pages}{418--438}
  (\bibinfo{year}{2010}).

\bibitem{Wang2017}
\bibinfo{author}{Wang, Y.}, \bibinfo{author}{Chen, X.}, \bibinfo{author}{Gao,
  X.} \& \bibinfo{author}{Gao, S.}
\newblock \bibinfo{title}{A benchmark dataset for {SSVEP}-based brain-computer
  interfaces.}
\newblock \emph{\bibinfo{journal}{IEEE Trans. on Neural Systems and
  Rehabilitation Engineering}} \textbf{\bibinfo{volume}{25}},
  \bibinfo{pages}{1746--1752} (\bibinfo{year}{2017}).

\bibitem{Szegedy2014}
\bibinfo{author}{Szegedy, C.} \emph{et~al.}
\newblock \bibinfo{title}{Intriguing properties of neural networks}.
\newblock In \emph{\bibinfo{booktitle}{Proc. Int'l Conf. on Learning
  Representations}} (\bibinfo{address}{Banff, Canada}, \bibinfo{year}{2014}).

\bibitem{Goodfellow2015}
\bibinfo{author}{Goodfellow, I.~J.}, \bibinfo{author}{Shlens, J.} \&
  \bibinfo{author}{Szegedy, C.}
\newblock \bibinfo{title}{Explaining and harnessing adversarial examples}.
\newblock In \emph{\bibinfo{booktitle}{Proc. Int'l Conf. on Learning
  Representations}} (\bibinfo{address}{San Diego, CA}, \bibinfo{year}{2015}).

\bibitem{Kurakin2017}
\bibinfo{author}{Kurakin, A.}, \bibinfo{author}{Goodfellow, I.~J.} \&
  \bibinfo{author}{Bengio, S.}
\newblock \bibinfo{title}{Adversarial examples in the physical world}.
\newblock In \emph{\bibinfo{booktitle}{Proc. Int'l Conf. on Learning
  Representations}} (\bibinfo{address}{Toulon, France}, \bibinfo{year}{2017}).

\bibitem{Athalye2018}
\bibinfo{author}{Athalye, A.}, \bibinfo{author}{Engstrom, L.},
  \bibinfo{author}{Ilyas, A.} \& \bibinfo{author}{Kwok, K.}
\newblock \bibinfo{title}{Synthesizing robust adversarial examples}.
\newblock In \emph{\bibinfo{booktitle}{Proc. 35th Int'l Conf. on Machine
  Learning}}, \bibinfo{pages}{284--293} (\bibinfo{address}{Stockholm, Sweden},
  \bibinfo{year}{2018}).

\bibitem{Papernot2016}
\bibinfo{author}{Papernot, N.}, \bibinfo{author}{McDaniel, P.} \&
  \bibinfo{author}{Goodfellow, I.}
\newblock \bibinfo{title}{Transferability in machine learning: {F}rom phenomena
  to black-box attacks using adversarial samples}.
\newblock \emph{\bibinfo{journal}{CoRR}}
  \textbf{\bibinfo{volume}{abs/1605.07277}} (\bibinfo{year}{2016}).
\newblock \urlprefix\url{https://arxiv.org/abs/1605.07277}.

\bibitem{Carlini2018}
\bibinfo{author}{Carlini, N.} \& \bibinfo{author}{Wagner, D.~A.}
\newblock \bibinfo{title}{Audio adversarial examples: {T}argeted attacks on
  speech-to-text}.
\newblock In \emph{\bibinfo{booktitle}{Proc. {IEEE} Symposium on Security and
  Privacy}}, \bibinfo{pages}{1--7} (\bibinfo{address}{San Francisco, CA},
  \bibinfo{year}{2018}).

\bibitem{Jia2017}
\bibinfo{author}{Jia, R.} \& \bibinfo{author}{Liang, P.}
\newblock \bibinfo{title}{Adversarial examples for evaluating reading
  comprehension systems}.
\newblock \emph{\bibinfo{journal}{CoRR}}
  \textbf{\bibinfo{volume}{abs/1707.07328}} (\bibinfo{year}{2017}).
\newblock \urlprefix\url{https://arxiv.org/abs/1707.07328}.

\bibitem{Grosse2016}
\bibinfo{author}{Grosse, K.}, \bibinfo{author}{Papernot, N.},
  \bibinfo{author}{Manoharan, P.}, \bibinfo{author}{Backes, M.} \&
  \bibinfo{author}{McDaniel, P.}
\newblock \bibinfo{title}{Adversarial perturbations against deep neural
  networks for malware classification}.
\newblock \emph{\bibinfo{journal}{CoRR}}
  \textbf{\bibinfo{volume}{abs/1606.04435}} (\bibinfo{year}{2016}).
\newblock \urlprefix\url{https://arxiv.org/abs/1606.04435}.

\bibitem{Papernot2016a}
\bibinfo{author}{Papernot, N.}, \bibinfo{author}{McDaniel, P.},
  \bibinfo{author}{Wu, X.}, \bibinfo{author}{Jha, S.} \&
  \bibinfo{author}{Swami, A.}
\newblock \bibinfo{title}{Distillation as a defense to adversarial
  perturbations against deep neural networks}.
\newblock In \emph{\bibinfo{booktitle}{Proc. {IEEE} Symposium on Security and
  Privacy}}, \bibinfo{pages}{582--597} (\bibinfo{address}{San Jose, CA},
  \bibinfo{year}{2016}).

\bibitem{Madry2018}
\bibinfo{author}{Madry, A.}, \bibinfo{author}{Makelov, A.},
  \bibinfo{author}{Schmidt, L.}, \bibinfo{author}{Tsipras, D.} \&
  \bibinfo{author}{Vladu, A.}
\newblock \bibinfo{title}{Towards deep learning models resistant to adversarial
  attacks}.
\newblock In \emph{\bibinfo{booktitle}{Proc. Int'l Conf. on Learning
  Representations}} (\bibinfo{address}{Vancouver, Canada},
  \bibinfo{year}{2018}).

\bibitem{Tramer2018}
\bibinfo{author}{Tramèr, F.} \emph{et~al.}
\newblock \bibinfo{title}{Ensemble adversarial training: {A}ttacks and
  defenses}.
\newblock In \emph{\bibinfo{booktitle}{Proc. Int'l Conf. on Learning
  Representations}} (\bibinfo{address}{Vancouver, Canada},
  \bibinfo{year}{2018}).

\bibitem{Samangouei2018}
\bibinfo{author}{Samangouei, P.}, \bibinfo{author}{Kabkab, M.} \&
  \bibinfo{author}{Chellappa, R.}
\newblock \bibinfo{title}{Defense-{GAN}: {P}rotecting classifiers against
  adversarial attacks using generative models}.
\newblock In \emph{\bibinfo{booktitle}{Proc. Int'l Conf. on Learning
  Representations}} (\bibinfo{address}{Vancouver, Canada},
  \bibinfo{year}{2018}).

\bibitem{Xie2018}
\bibinfo{author}{Xie, C.}, \bibinfo{author}{Wang, J.}, \bibinfo{author}{Zhang,
  Z.}, \bibinfo{author}{Ren, Z.} \& \bibinfo{author}{Yuille, A.}
\newblock \bibinfo{title}{Mitigating adversarial effects through
  randomization}.
\newblock In \emph{\bibinfo{booktitle}{Proc. Int'l Conf. on Learning
  Representations}} (\bibinfo{address}{Vancouver, Canada},
  \bibinfo{year}{2018}).

\bibitem{Qin2019}
\bibinfo{author}{Qin, C.} \emph{et~al.}
\newblock \bibinfo{title}{Adversarial robustness through local linearization}.
\newblock In \emph{\bibinfo{booktitle}{Proc. Advances in Neural Information
  Processing Systems}}, \bibinfo{pages}{13824--13833}
  (\bibinfo{address}{Vancouver, Canada}, \bibinfo{year}{2019}).

\bibitem{Carlini2017}
\bibinfo{author}{Carlini, N.} \& \bibinfo{author}{Wagner, D.}
\newblock \bibinfo{title}{Towards evaluating the robustness of neural
  networks}.
\newblock In \emph{\bibinfo{booktitle}{Proc. {IEEE} Symposium on Security and
  Privacy}}, \bibinfo{pages}{39--57} (\bibinfo{address}{San Jose, CA},
  \bibinfo{year}{2017}).

\bibitem{Athalye2018a}
\bibinfo{author}{Athalye, A.}, \bibinfo{author}{Carlini, N.} \&
  \bibinfo{author}{Wagner, D.}
\newblock \bibinfo{title}{Obfuscated gradients give a false sense of security:
  {C}ircumventing defenses to adversarial examples}.
\newblock In \emph{\bibinfo{booktitle}{Proc. 35th Int'l Conf. on Machine
  Learning}}, \bibinfo{pages}{274--283} (\bibinfo{address}{Stockholm, Sweden},
  \bibinfo{year}{2018}).

\bibitem{Gowal2018}
\bibinfo{author}{Gowal, S.} \emph{et~al.}
\newblock \bibinfo{title}{On the effectiveness of interval bound propagation
  for training verifiably robust models}.
\newblock \emph{\bibinfo{journal}{CoRR}}
  \textbf{\bibinfo{volume}{abs/1810.12715}} (\bibinfo{year}{2018}).
\newblock \urlprefix\url{http://arXiv.org/abs/1810.12715}.

\bibitem{Cohen2019}
\bibinfo{author}{Cohen, J.}, \bibinfo{author}{Rosenfeld, E.} \&
  \bibinfo{author}{Kolter, Z.}
\newblock \bibinfo{title}{Certified adversarial robustness via randomized
  smoothing}.
\newblock In \emph{\bibinfo{booktitle}{Proc. 36th Int'l Conf. on Machine
  Learning}}, \bibinfo{pages}{1310--1320} (\bibinfo{address}{Long Beach, CA},
  \bibinfo{year}{2019}).

\bibitem{Li2019}
\bibinfo{author}{Li, B.}, \bibinfo{author}{Chen, C.}, \bibinfo{author}{Wang,
  W.} \& \bibinfo{author}{Carin, L.}
\newblock \bibinfo{title}{Certified adversarial robustness with additive
  noise}.
\newblock In \emph{\bibinfo{booktitle}{Proc. Advances in Neural Information
  Processing Systems}}, \bibinfo{pages}{9464--9474}
  (\bibinfo{address}{Vancouver, Canada}, \bibinfo{year}{2019}).

\bibitem{Balunovic2020}
\bibinfo{author}{Balunovic, M.} \& \bibinfo{author}{Vechev, M.}
\newblock \bibinfo{title}{Adversarial training and provable defenses: Bridging
  the gap}.
\newblock In \emph{\bibinfo{booktitle}{Proc. Int'l Conf. on Learning
  Representations}} (\bibinfo{address}{Addis Ababa, Ethiopia},
  \bibinfo{year}{2020}).

\bibitem{Zhang2019}
\bibinfo{author}{Zhang, X.} \& \bibinfo{author}{Wu, D.}
\newblock \bibinfo{title}{On the vulnerability of {CNN} classifiers in
  {EEG}-based {BCI}s}.
\newblock \emph{\bibinfo{journal}{IEEE Trans. on Neural Systems and
  Rehabilitation Engineering}} \textbf{\bibinfo{volume}{27}},
  \bibinfo{pages}{814--825} (\bibinfo{year}{2019}).

\bibitem{Fawaz2019}
\bibinfo{author}{Fawaz, H.~I.}, \bibinfo{author}{Forestier, G.},
  \bibinfo{author}{Weber, J.}, \bibinfo{author}{Idoumghar, L.} \&
  \bibinfo{author}{Muller, P.-A.}
\newblock \bibinfo{title}{Adversarial attacks on deep neural networks for time
  series classification}.
\newblock In \emph{\bibinfo{booktitle}{Int'l Joint Conf. on Neural Networks}},
  \bibinfo{pages}{1--8} (\bibinfo{address}{Budapest, Hungary},
  \bibinfo{year}{2019}).

\bibitem{Qin2019a}
\bibinfo{author}{Qin, Y.}, \bibinfo{author}{Carlini, N.},
  \bibinfo{author}{Cottrell, G.}, \bibinfo{author}{Goodfellow, I.} \&
  \bibinfo{author}{Raffel, C.}
\newblock \bibinfo{title}{Imperceptible, robust, and targeted adversarial
  examples for automatic speech recognition}.
\newblock In \emph{\bibinfo{booktitle}{Proc. 36th Int'l Conf. on Machine
  Learning}}, \bibinfo{pages}{5231--5240} (\bibinfo{address}{Long Beach, CA},
  \bibinfo{year}{2019}).

\bibitem{Moosavi-Dezfooli2017}
\bibinfo{author}{Moosavi-Dezfooli, S.-M.}, \bibinfo{author}{Fawzi, A.},
  \bibinfo{author}{Fawzi, O.} \& \bibinfo{author}{Frossard, P.}
\newblock \bibinfo{title}{Universal adversarial perturbations}.
\newblock In \emph{\bibinfo{booktitle}{Proc. IEEE Conf. on Computer Vision and
  Pattern Recognition}}, \bibinfo{pages}{1765--1773}
  (\bibinfo{address}{Honolulu, HI}, \bibinfo{year}{2017}).

\bibitem{Wolpaw1998}
\bibinfo{author}{Wolpaw, J.~R.}, \bibinfo{author}{Ramoser, H.},
  \bibinfo{author}{McFarland, D.~J.} \& \bibinfo{author}{Pfurtscheller, G.}
\newblock \bibinfo{title}{{EEG}-based communication: {I}mproved accuracy by
  response verification}.
\newblock \emph{\bibinfo{journal}{IEEE Trans. on Rehabilitation Engineering}}
  \textbf{\bibinfo{volume}{6}}, \bibinfo{pages}{326--333}
  (\bibinfo{year}{1998}).

\bibitem{Wolpaw2004}
\bibinfo{author}{Wolpaw, J.~R.}, \bibinfo{author}{Krusienski, D.} \&
  \bibinfo{author}{Schalk, G.}
\newblock \bibinfo{title}{Documentation {Wadsworth} {BCI} dataset ({P}300
  evoked potentials)} (\bibinfo{year}{2004}).
\newblock \urlprefix\url{http://www.bbci.de/competition/iii/desc_II.pdf}.

\bibitem{Rivet2009}
\bibinfo{author}{Rivet, B.}, \bibinfo{author}{Souloumiac, A.},
  \bibinfo{author}{Attina, V.} \& \bibinfo{author}{Gibert, G.}
\newblock \bibinfo{title}{x{DAWN} algorithm to enhance evoked potentials:
  {A}pplication to brain-computer interface}.
\newblock \emph{\bibinfo{journal}{IEEE Trans. on Biomedical Engineering}}
  \textbf{\bibinfo{volume}{56}}, \bibinfo{pages}{2035--2043}
  (\bibinfo{year}{2009}).

\bibitem{Barachant2012}
\bibinfo{author}{Barachant, A.}, \bibinfo{author}{Bonnet, S.},
  \bibinfo{author}{Congedo, M.} \& \bibinfo{author}{Jutten, C.}
\newblock \bibinfo{title}{Multiclass brain-computer interface classification by
  {R}iemannian geometry}.
\newblock \emph{\bibinfo{journal}{IEEE Trans. on Biomedical Engineering}}
  \textbf{\bibinfo{volume}{59}}, \bibinfo{pages}{920--928}
  (\bibinfo{year}{2012}).

\bibitem{Barachant2013}
\bibinfo{author}{Barachant, A.}, \bibinfo{author}{Bonnet, S.},
  \bibinfo{author}{Congedo, M.} \& \bibinfo{author}{Jutten, C.}
\newblock \bibinfo{title}{Classification of covariance matrices using a
  {R}iemannian-based kernel for {BCI} applications}.
\newblock \emph{\bibinfo{journal}{Neurocomputing}}
  \textbf{\bibinfo{volume}{112}}, \bibinfo{pages}{172--178}
  (\bibinfo{year}{2013}).

\bibitem{Yger2017}
\bibinfo{author}{Yger, F.}, \bibinfo{author}{Berar, M.} \&
  \bibinfo{author}{Lotte, F.}
\newblock \bibinfo{title}{{R}iemannian approaches in brain-computer interfaces:
  {A} review}.
\newblock \emph{\bibinfo{journal}{IEEE Trans. on Neural Systems and
  Rehabilitation Engineering}} \textbf{\bibinfo{volume}{25}},
  \bibinfo{pages}{1753--1762} (\bibinfo{year}{2017}).

\bibitem{Abadi2016}
\bibinfo{author}{Abadi, M.} \emph{et~al.}
\newblock \bibinfo{title}{Tensorflow: {A} system for large-scale machine
  learning}.
\newblock In \emph{\bibinfo{booktitle}{Proc. 12th {USENIX} Symposium on
  Operating Systems Design and Implementation}}, \bibinfo{pages}{265--283}
  (\bibinfo{address}{Savannah, GA}, \bibinfo{year}{2016}).

\bibitem{Chen2015}
\bibinfo{author}{Chen, X.}, \bibinfo{author}{Wang, Y.}, \bibinfo{author}{Gao,
  S.}, \bibinfo{author}{Jung, T.-P.} \& \bibinfo{author}{Gao, X.}
\newblock \bibinfo{title}{Filter bank canonical correlation analysis for
  implementing a high-speed {SSVEP}-based brain--computer interface}.
\newblock \emph{\bibinfo{journal}{Journal of Neural Engineering}}
  \textbf{\bibinfo{volume}{12}}, \bibinfo{pages}{046008}
  (\bibinfo{year}{2015}).

\bibitem{Lin2006}
\bibinfo{author}{Lin, Z.}, \bibinfo{author}{Zhang, C.}, \bibinfo{author}{Wu,
  W.} \& \bibinfo{author}{Gao, X.}
\newblock \bibinfo{title}{Frequency recognition based on canonical correlation
  analysis for {SSVEP}-based {BCI}s}.
\newblock \emph{\bibinfo{journal}{IEEE Trans. on Biomedical Engineering}}
  \textbf{\bibinfo{volume}{53}}, \bibinfo{pages}{2610--2614}
  (\bibinfo{year}{2006}).

\bibitem{Akaike1976}
\bibinfo{author}{Akaike, H.}
\newblock \bibinfo{title}{Canonical correlation analysis of time series and the
  use of an information criterion}.
\newblock In \emph{\bibinfo{booktitle}{Mathematics in Science and
  Engineering}}, vol. \bibinfo{volume}{126}, \bibinfo{pages}{27--96}
  (\bibinfo{publisher}{Elsevier}, \bibinfo{year}{1976}).

\bibitem{Liu2016}
\bibinfo{author}{Liu, Y.}, \bibinfo{author}{Chen, X.}, \bibinfo{author}{Liu,
  C.} \& \bibinfo{author}{Song, D.}
\newblock \bibinfo{title}{Delving into transferable adversarial examples and
  black-box attacks}.
\newblock \emph{\bibinfo{journal}{CoRR}}
  \textbf{\bibinfo{volume}{abs/1611.02770}} (\bibinfo{year}{2016}).
\newblock \urlprefix\url{http://arxiv.org/abs/1611.02770}.

\bibitem{Tramer2017}
\bibinfo{author}{Tramèr, F.}, \bibinfo{author}{Papernot, N.},
  \bibinfo{author}{Goodfellow, I.}, \bibinfo{author}{Boneh, D.} \&
  \bibinfo{author}{McDaniel, P.}
\newblock \bibinfo{title}{The space of transferable adversarial examples}.
\newblock \emph{\bibinfo{journal}{CoRR}}
  \textbf{\bibinfo{volume}{abs/1704.03453v2}} (\bibinfo{year}{2017}).
\newblock \urlprefix\url{http://arXiv.org/abs/1704.03453v2}.

\bibitem{Wu2018}
\bibinfo{author}{Wu, L.}, \bibinfo{author}{Zhu, Z.}, \bibinfo{author}{Tai, C.}
  \& \bibinfo{author}{E, W.}
\newblock \bibinfo{title}{Understanding and enhancing the transferability of
  adversarial examples}.
\newblock \emph{\bibinfo{journal}{CoRR}}
  \textbf{\bibinfo{volume}{abs/1802.09707}} (\bibinfo{year}{2018}).
\newblock \urlprefix\url{http://arXiv.org/abs/1802.09707}.

\bibitem{Brown2017}
\bibinfo{author}{Brown, T.~B.}, \bibinfo{author}{Man{\'{e}}, D.},
  \bibinfo{author}{Roy, A.}, \bibinfo{author}{Abadi, M.} \&
  \bibinfo{author}{Gilmer, J.}
\newblock \bibinfo{title}{Adversarial patch}.
\newblock \emph{\bibinfo{journal}{CoRR}}
  \textbf{\bibinfo{volume}{abs/1712.09665}} (\bibinfo{year}{2017}).
\newblock \urlprefix\url{http://arxiv.org/abs/1712.09665}.

\bibitem{Li2016}
\bibinfo{author}{Li, Y.} \emph{et~al.}
\newblock \bibinfo{title}{Multimodal {BCIs}: {T}arget detection,
  multidimensional control, and awareness evaluation in patients with disorder
  of consciousness}.
\newblock \emph{\bibinfo{journal}{Proceedings of the {IEEE}}}
  \textbf{\bibinfo{volume}{104}}, \bibinfo{pages}{332--352}
  (\bibinfo{year}{2016}).

\bibitem{Riccio2013}
\bibinfo{author}{Riccio, A.} \emph{et~al.}
\newblock \bibinfo{title}{Attention and {P}300-based {BCI} performance in
  people with amyotrophic lateral sclerosis}.
\newblock \emph{\bibinfo{journal}{Frontiers in Human Neuroscience}}
  \textbf{\bibinfo{volume}{7}}, \bibinfo{pages}{732} (\bibinfo{year}{2013}).

\bibitem{Hestenes1969}
\bibinfo{author}{Hestenes, M.~R.}
\newblock \bibinfo{title}{Multiplier and gradient methods}.
\newblock \emph{\bibinfo{journal}{Journal of Optimization Theory and
  Applications}} \textbf{\bibinfo{volume}{4}}, \bibinfo{pages}{303--320}
  (\bibinfo{year}{1969}).

\end{thebibliography}
\end{document}